\documentclass[titlepage]{article}

\usepackage{amsmath}
\usepackage{amssymb}

\if x\pdfoutput\undefined   
\usepackage[dvips]{graphicx}   
\else   
\usepackage[pdftex]{graphicx}
\fi
\usepackage{wrapfig}
\usepackage{lscape}
\usepackage{rotating}
\usepackage{epstopdf}
\usepackage{subfigure,amsmath,amsfonts}
\graphicspath{{./},{./graphics/}}

\usepackage{wrapfig}

\allowdisplaybreaks

\usepackage[round,sort&compress]{natbib}
\newcommand*\diff{\mathop{}\!\mathrm{d}}

\title{\bf Deformation of a Half-Space from Anelastic Strain Confined in a Tetrahedral Volume}
\author{Sylvain Barbot$^{1}$\\$^1$Earth Observatory of Singapore, Nanyang Technological University}
\date{}

\makeatletter
\let\ps@oldempty\ps@empty 
\renewcommand\ps@empty\ps@plain
\makeatother

\begin{document}

\maketitle

\section*{Abstract}
Deformation in the lithosphere-asthenosphere system can be accommodated by faulting and plastic flow. However, incorporating structural data in models of distributed deformation still represents a challenge. Here, I present solutions for the displacements and stress in a half-space caused by distributed anelastic strain confined in a tetrahedral volume. These solutions form the basis of curvilinear meshes that can adapt to realistic structural settings, such as a mantle wedge corner, a spherical shell around a magma chamber, or an aquifer. I provide computer programs to evaluate them in the cases of anti-plane strain, in-plane strain, and three-dimensional deformation. These tools may prove useful in the modeling of deformation data in tectonics, volcanology, and hydrology.

\section*{Introduction}

Earth's deformation encompasses physical processes that spread widely across space-time. The deformation of the lithosphere-asthenosphere system is largely accommodated by localized (faulting) and distributed (e.g., plastic flow, multi-phase flow) deformation. Because of the urgency of understanding seismic hazards, a large body of work is dedicated to describing brittle deformation~\citep{steketee58b,chinnery63,savage+hastie66,sato+matsuura74,iwasaki+sato79,jeyakumaran+92,wang+03a,okada85,okada92,meade07a,nikkhoo+15}. Recently,~\cite{barbot+17} described how distributed plastic deformation induces displacement and stress in the surrounding medium, opening the door to low-frequency and time-dependent tomography from deformation data~\citep{tsang+16,moore+17,qqiu+18} and to more comprehensive forward models of deformation in the lithosphere-asthenosphere system that include the mechanical coupling between brittle and viscoelastic deformation~\citep{lambert+barbot16,barbot18b}. 

\begin{figure}
\includegraphics[width=\columnwidth]{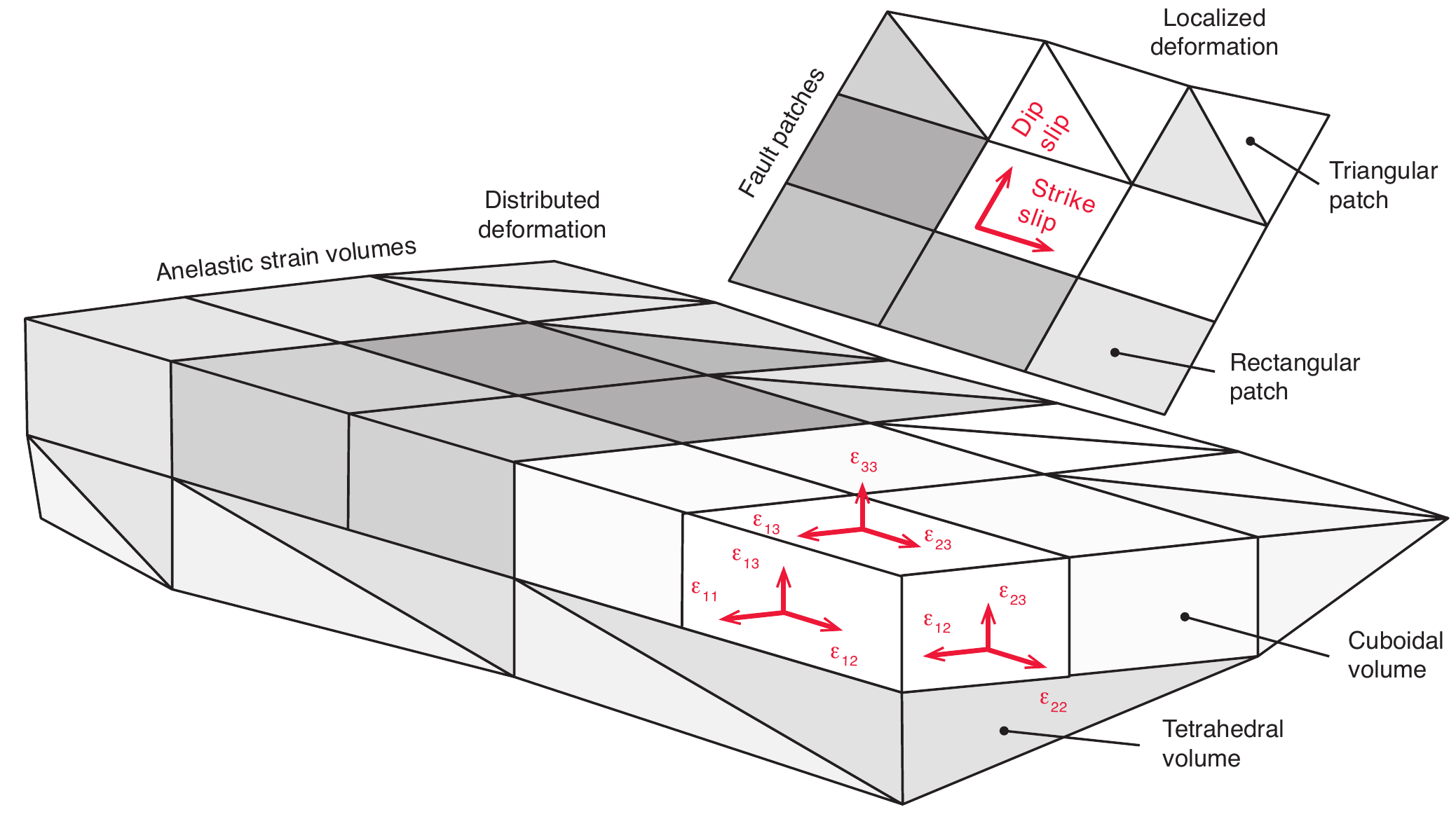}
\caption{Schematic view of the modeling approach. Localized deformation is discretized with triangular or rectangular boundary elements representing fault slip. Distributed deformation is discretized with tetrahedral or cuboidal volume elements representing plastic deformation. The surrounding elastic material is not meshed but its effect is included in the Green's functions. Curvilinear surfaces and volumes can be approximated with triangular and tetrahedral elements.}\label{fig:unified-discretization-cuboid-tetrahedra}
\end{figure}

Increasingly accurate images of Earth's internal strain and strain-rates require incorporating morphological gradients~\citep[e.g.,][]{murray+langbein06,walter+amelung06,marshall+09,dieterich+richards-dinger10,barnhart+lohman10,furuya+yasuda11,steer+14,li+liu16,qqiu+16}. A familiar approach in fault mechanics is to discretize faults in triangular elements, as they can conform to curvilinear surfaces at least to first-order approximation~\citep{yoffe60,comninou+dundurs75,jeyakumaran+92,gosling+willis94,maerten+05,meade07a,nikkhoo+15,ohtani+hirahara15}. It is natural to extend the approach to tetrahedral volumes for distributed anelastic strain to conform volume meshes to structural data. Rectangular and triangular fault elements and cuboidal and tetrahedral volumes can be combined to represent various physical processes of deformation in a realistic geometry. Figure~\ref{fig:unified-discretization-cuboid-tetrahedra} illustrates how different types of fault and volume elements can be combined to represent the kinematics or quasi-dynamics of a regional block of the lithosphere-asthenosphere system. Fault processes can be represented by triangular or rectangular boundary elements and distributed deformation processes can be discretized with tetrahedral or cuboidal volume elements.

In this paper, I focus on anelastic deformation confined in a tetrahedral volume for three-dimensional problems and triangular surfaces for two-dimensional problems. More complex deformation can be reproduced by a linear combination of these elementary solutions. In the next two sections, I describe the governing equations and derive a simple general expression for the displacement kernels for arbitrary volumes of quasi-static anelastic deformation. Then, I derive the displacement and stress kernels for the cases of anti-plane strain, plane strain, and three-dimensional deformation. In the last section, I derive numerical solutions based on fast Fourier transforms that are more amenable to large-scale problems. 

\section*{Eigenstrain and equivalent body forces}

The deformation of materials can be broadly categorized into elastic and anelastic deformation. Elastic deformation is reversible, implying that the material spontaneously recovers its original configuration when the loads are removed. Until then, the material remains under stress, following the constitutive stress/elastic strain relationship
\begin{equation}\label{eqn:constitutive}
\boldsymbol{\sigma}=\textbf{C}:\boldsymbol{\epsilon}^e~,
\end{equation}
where $\boldsymbol{\sigma}$ is the Cauchy stress, $\textbf{C}$ is the elastic moduli tensor, assumed independent of anelastic strain, and $\boldsymbol{\epsilon}^e$ is the elastic strain tensor. Anelastic deformation requires additional work to place the material back into its original configuration and is thermodynamically irreversible. Many deformation processes within the Earth, such as poroelasticity, viscoelasticity, and faulting are anelastic~\citep{barbot+fialko10a,Barbot+Fialko10b}. Therefore, manipulating the total strain in the medium as the sum of the elastic and anelastic contributions~\citep[e.g.,][]{andrews78}
\begin{equation}\label{eqn:total-strain}
\boldsymbol{\epsilon}=\boldsymbol{\epsilon}^e+\boldsymbol{\epsilon}^i~,
\end{equation}
where $\boldsymbol{\epsilon}^i$ represents the cumulative anelastic strain, is a useful approximation. In practical applications the anelastic strain or its time derivative is known, either provided by the constitutive behavior of the material under a given stress~\citep[e.g.,][]{barbot18b} or inverted for~\citep[e.g.,][]{qqiu+18}. The conservation of linear momentum at steady state leads to the following governing equation for the total strain
\begin{equation}\label{eqn:momentum}
\nabla\cdot\left(\textbf{C}:\boldsymbol{\epsilon}\right)+\textbf{f}=\textbf{0}~,
\end{equation}
where the anelastic strain has been associated with the equivalent body-force density
\begin{equation}\label{eqn:eqbf}
\textbf{f}=-\nabla\cdot\textbf{m}~,
\end{equation}
and the moment density $\textbf{m}=\textbf{C}:\boldsymbol{\epsilon}^i$. The total displacement $\textbf{u}(\textbf{x})$ due to anelastic strain and the elastic response of the medium can be obtained by solving the momentum equation (\ref{eqn:momentum}). The total strain follows as
\begin{equation}\label{eqn:displacement-strain}
\boldsymbol{\epsilon}=\frac{1}{2}\left(\nabla\textbf{u}+\nabla\textbf{u}^t\right)~,
\end{equation}
where $\nabla\textbf{u}^t$ is the transpose of the displacement gradient. Finally, the stress field is derived by removing the anelastic strain contribution combining (\ref{eqn:constitutive}) and (\ref{eqn:total-strain}), as follows
\begin{equation}\label{eqn:stress-eigenstrain}
\boldsymbol{\sigma}=\textbf{C}:\left(\boldsymbol{\epsilon}-\boldsymbol{\epsilon}^i\right)~.
\end{equation}
Navier's equation (\ref{eqn:momentum}) applies to quasi-static deformation due to an arbitrary distribution of anelastic strain under the infinitesimal strain approximation and is valid as long as plastic deformation does not affect the elastic moduli in the medium and inertia can be ignored. As a numerical approximation, I assume piecewise uniform anelastic strain distributions, called transformation strain, within closed volumes $\Omega_k$, so that the displacement can be written as
\begin{equation}\label{eqn:mesh}
\textbf{u}(\textbf{x})\approx \sum_k \int_{\Omega_k} \textbf{G}(\textbf{x},\textbf{y})\cdot\textbf{f}_k(\textbf{y}) \,\diff \textbf{y}~,
\end{equation}
where $\textbf{f}_k$ is the equivalent body force for a homogeneous anelastic strain in the domain $\Omega_k$ and $\textbf{G}(\textbf{x};\textbf{y})$ are the Green's functions for a point force. The displacement kernels in (\ref{eqn:mesh}) form the basic ingredients for forward~\citep{lambert+barbot16,barbot18b} and inverse~\citep{tsang+16,moore+17,qqiu+18} modeling of deformation. The closed-form analytic solution of (\ref{eqn:mesh}) for cuboid volumes of transformation strain is provided by~\cite{barbot+17}. To facilitate the meshing of curvilinear surfaces and volumes, I now make the assumption that the transformation strain is confined in a tetrahedral volume.

\section*{Displacement kernels}

To develop the solution for the displacement kernel (I drop the subscript $k$ for the sake of clarity)
\begin{equation}\label{eqn:kernel}
\textbf{u}(\textbf{x})= \int_{\Omega} \textbf{G}(\textbf{x},\textbf{y})\cdot\textbf{f}(\textbf{y}) \,\diff \textbf{y}
\end{equation}
associated with a uniform transformation strain confined within a elementary volume $\Omega$, I write the moment density as
\begin{equation}
\textbf{m}(\textbf{x})=\Phi(\textbf{x})\,\textbf{m}_0~,
\end{equation}
where $\Phi(\textbf{x})$ is a single-variate function that represents the location of the transformation strain,
\begin{equation}
\Phi(\textbf{x})=\left\{\begin{aligned}
1& & &\text{if }\textbf{x}\in\Omega, \\
0& & &\text{otherwise~,}
\end{aligned}\right.
\end{equation}
and $\textbf{m}_0$ is a constant tensor. With this definition, the equivalent body force becomes
\begin{equation}\label{eqn:eqbf-simplified}
\textbf{f}=-\textbf{m}_0\cdot\nabla\Phi~.
\end{equation}
As $\Phi(\textbf{x})$ is uniform within $\Omega$, I can write
\begin{equation}\label{eqn:grad-pi}
\nabla\Phi=\left\{\begin{aligned}
-\textbf{n}& & &\text{if }\textbf{x}\in\partial\Omega \\
0& & &\text{otherwise,}
\end{aligned}\right.
\end{equation}
where $\textbf{n}$ is the outward-pointing unit normal vector to $\Omega$. Combining (\ref{eqn:kernel}), (\ref{eqn:eqbf-simplified}), and (\ref{eqn:grad-pi}), the displacement kernel simplifies to the surface integral
\begin{equation}\label{eqn:kernel-simplified}
\begin{aligned}
\textbf{u}(\textbf{x})&=\textbf{m}_0\cdot\int_{\partial\Omega} \textbf{G}(\textbf{x},\textbf{y})\cdot\textbf{n}(\textbf{y}) \,\diff \textbf{y}~. \\
\end{aligned}
\end{equation}
The stress can be obtained by differentiation of the Green's function $\textbf{G}(\textbf{x},\textbf{y})$ itself or of the resulting displacement field, following (\ref{eqn:displacement-strain}) and (\ref{eqn:stress-eigenstrain}). Equation (\ref{eqn:kernel-simplified}) represents a convenient framework to evaluate the deformation due to transformation strain confined in volumes of arbitrary shape as the integral equation simplifies to a path integral in two dimensions or to a surface integral in three dimension, whereas the form (\ref{eqn:kernel}) requires a surface integral in two dimensions and a volume integral in three dimensions. In the next sections, I develop solutions for these kernels for triangular surfaces in the cases of anti-plane strain and plane strain and for tetrahedral volumes in the case of three-dimensional deformation. The solution for more complex shapes can be obtained by superposition using the approximation (\ref{eqn:mesh}).

\section*{Distributed deformation of triangular shear zones in anti-plane strain}

Two-dimensional models of stress evolution may capture the main features of a mechanical setting~\citep{savage+prescott78,thatcher+rundle79,savage83} and their reduced complexity is more amenable to sensitivity analyses~\citep{daout+16a,daout+16b,muto+16}. The anti-plane strain approximation is relevant to transform plate boundaries~\citep[e.g.,][]{nur+mavko74a,nur+israel80,barbot+08b,lindsey+14,lambert+barbot16,erickson+17} and curvilinear elements may represent shear zones~\citep[e.g.,][]{takeuchi+fialko13} or lower-crustal flow within a realistic stratigraphy.

\begin{figure}
\includegraphics[width=\columnwidth]{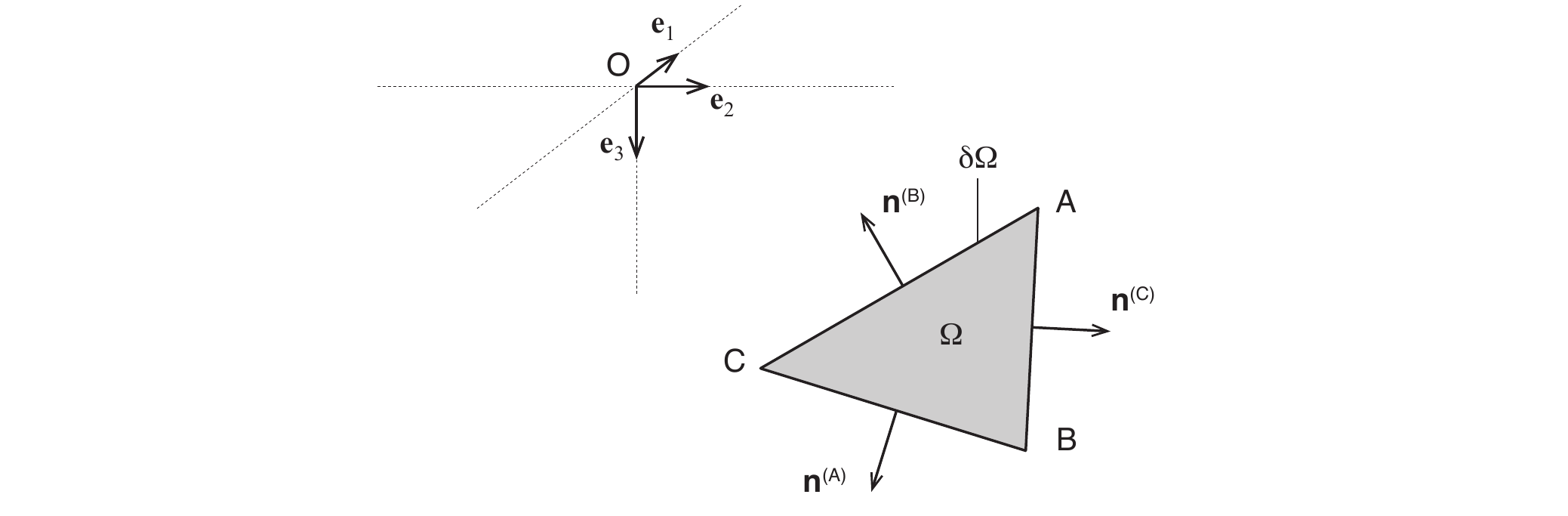}
\caption{Deformation of a half-space in anti-plane strain due to anelastic strain confined in a triangular element ABC. The vertices A, B, and C have the coordinates $\textbf{x}^A$, $\textbf{x}^B$, and $\textbf{x}^C$, respectively. The normal vectors are pointing outwards, such that $\textbf{n}^{(C)}\cdot(\textbf{x}_C-\textbf{x}_A)\le0$ and $\textbf{n}^{(C)}\cdot(\textbf{x}_C-\textbf{x}_B)\le0$.}\label{fig:geometry-antiplane-triangle}
\end{figure}

\subsection*{Problem statement}

Consider the elastic deformation in a half-space of rigidity $\mu$ in a situation of anti-plane strain caused by distributed anelastic strain confined in an elementary triangular area. In the case of anti-plane strain we have $u_{i,1}=0$ for $i=1,2,3$ and $u_2=u_3=0$. The transformation strain is confined in a triangular area delimited by three points A, B, and C (Figure~\ref{fig:geometry-antiplane-triangle}). The surface is subjected to two independent transformation strain components $\epsilon^i_{12}$ and $\epsilon^i_{13}$ associated with the moment density $m_{12}=2\mu\epsilon_{12}^i$ and $m_{13}=2\mu\epsilon_{13}^i$. Using (\ref{eqn:kernel-simplified}), the deformation simplifies to the nontrivial component
\begin{equation}
u_1(x_2,x_3)=\int_{\partial\Omega}G_{11}(x_2,x_3,y_2,y_3)\left(m_{12}n_2+m_{13}n_3\right)\,\diff y_2\diff y_3~,
\end{equation}
where the Green's function for a line force centered at $(y_2,y_3)$ is obtained by solving Poisson's equation with a Neumann boundary condition and is given by
\begin{equation}\label{eqn:greens-function-1d}
G_{11}(x_2,x_3)=-\frac{1}{4\pi \mu}\bigg[\ln\left((x_2-y_2)^2+(x_3-y_3)^2\right) + \ln\left((x_2-y_2)^2+(x_3+y_3)^2\right)\bigg]~.
\end{equation}
The outward normal vector is different on each side, so we can write
\begin{equation}\label{eqn:line-integrals}
\begin{aligned}
u_1(x_2,x_3)&=\left(m_{12}n_2^{(C)}+m_{13}n_3^{(C)}\right)\int_{AB}G_{11}(x_2,x_3,y_2,y_3)\,\diff y_2\diff y_3 \\
&+\left(m_{12}n_2^{(A)}+m_{13}n_3^{(A)}\right)\int_{BC}G_{11}(x_2,x_3,y_2,y_3)\,\diff y_2\diff y_3 \\
&+\left(m_{12}n_2^{(B)}+m_{13}n_3^{(B)}\right)\int_{AC}G_{11}(x_2,x_3,y_2,y_3)\,\diff y_2\diff y_3~, \\
\end{aligned}
\end{equation}
where $\textbf{n}^{(A)}$, $\textbf{n}^{(B)}$, and $\textbf{n}^{(C)}$ are the unit normal vectors to the sides BC, AC, and AB, respectively. 

\subsection*{Analytic solution}

\begin{figure}[t]
\includegraphics[width=\columnwidth]{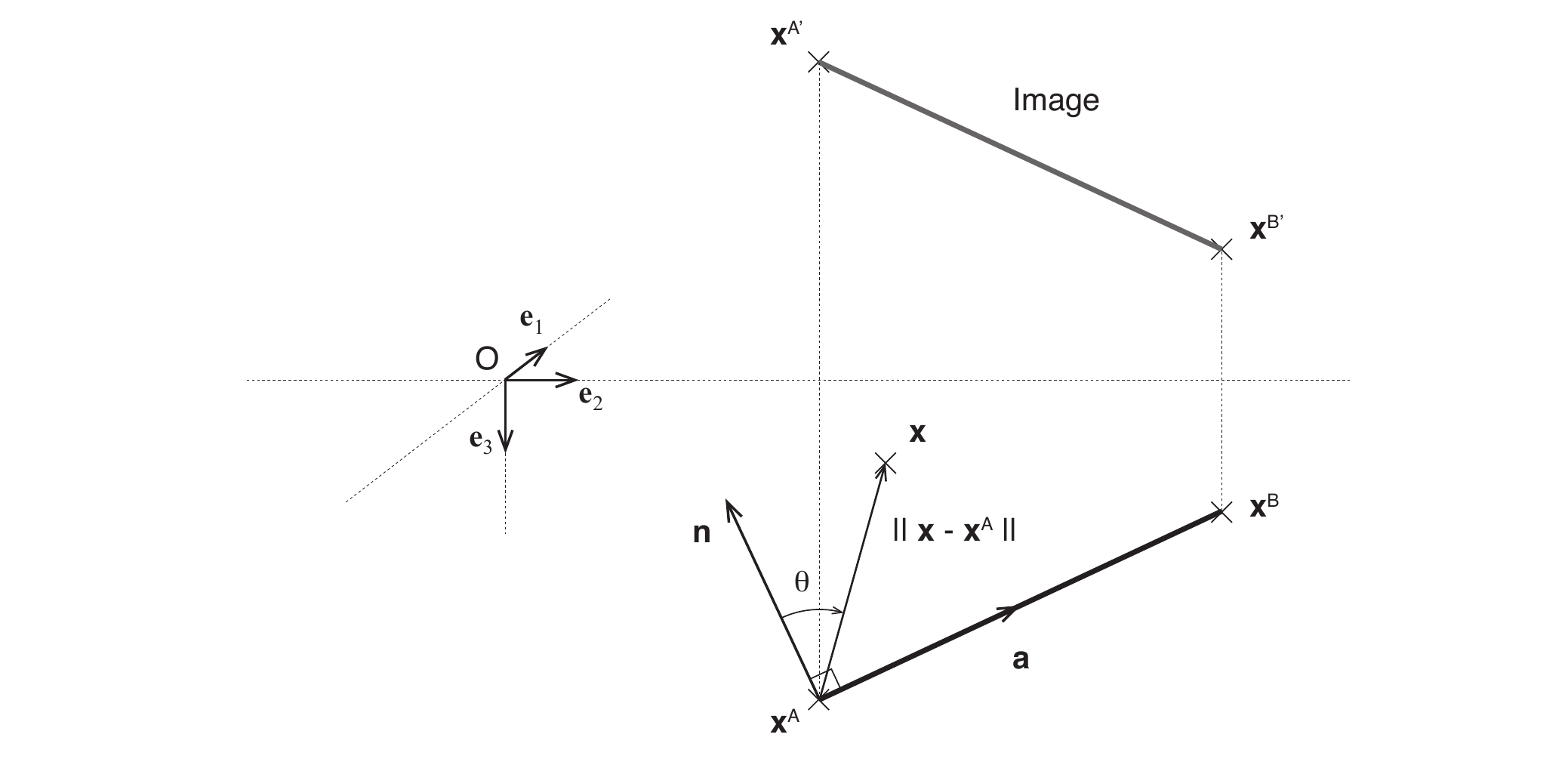}
\caption{The line integration of the Green's function $G_{11}$ along the segment AB only depend on the coordinates of the end-points $\textbf{x}^A$ and $\textbf{x}^B$. The unit vector $\textbf{a}$ is parallel to AB and the unit normal vector $\textbf{v}$ is perpendicular to AB. The image points $A'$ and $B'$ are defined to satisfy the free-surface boundary condition. The local angle is given by $\tan\theta=\textbf{a}\cdot(\textbf{x}-\textbf{r})\,/\,\textbf{n}\cdot(\textbf{x}-\textbf{r})$.}\label{fig:geometry-line-integral}
\end{figure}

The line integrals (\ref{eqn:line-integrals}) are path independent and only depend on the coordinates of the end-points. For any end point A and B, the closed-form solutions can be found using
\begin{equation}\label{eqn:closed-form-antiplane}
\begin{aligned}
\int_{AB}G_{11}(x_2,x_3,y_2,y_3)\,\diff y_2\diff y_3&=\Gamma(\textbf{x}^B)-\Gamma(\textbf{x}^A) \\
&+\Gamma(\textbf{x}^{B'})-\Gamma(\textbf{x}^{A'})~,
\end{aligned}
\end{equation}
where $\textbf{x}^A$ and $\textbf{x}^B$ are the coordinates of points A and B (Figure~\ref{fig:geometry-line-integral}), A' and B' are the images of points A and B about the surface, and $\Gamma(\textbf{r})$ is given by
\begin{equation}\label{eqn:gamma}
\begin{aligned}
\Gamma(\textbf{r})&=\frac{1}{8\pi}\textbf{a}\cdot(\textbf{x}-\textbf{r})\ln\left((\textbf{x}-\textbf{r})\cdot(\textbf{x}-\textbf{r})\right) \\
&\qquad\qquad+\frac{1}{4\pi}\,\textbf{n}\cdot(\textbf{x}-\textbf{r})\,\mathrm{arctan}\left[\frac{\textbf{a}\cdot(\textbf{x}-\textbf{r})}{\textbf{n}\cdot(\textbf{x}-\textbf{r})}\right]~,
\end{aligned}
\end{equation}
with the unit vector $\textbf{a}$ aligned with the segment AB and $\textbf{n}$ a unit vector normal to the segment AB, such that $\textbf{n}\cdot\textbf{a}=0$, and where I have removed the terms that cancel out upon integration over a closed path. The solution to (\ref{eqn:line-integrals}) is found by evaluating (\ref{eqn:closed-form-antiplane}) once for each segments and multiplying the result by the respective tractions. The displacement gradient is obtained in a similar way using
\begin{equation}\label{eqn:anti-plane-strain-analytic}
\begin{aligned}
\nabla\Gamma(\textbf{r})&=\frac{1}{8\pi}\textbf{a}\,\ln\left((\textbf{x}-\textbf{r})\cdot(\textbf{x}-\textbf{r})\right) \\
&\qquad\qquad+\frac{1}{4\pi}\,\textbf{n}\,\mathrm{arctan}\left[\frac{\textbf{a}\cdot(\textbf{x}-\textbf{r})}{\textbf{n}\cdot(\textbf{x}-\textbf{r})}\right]~,
\end{aligned}
\end{equation}
where I have again removed the terms that cancel out upon integration over a closed path. The expressions (\ref{eqn:gamma}) and (\ref{eqn:anti-plane-strain-analytic}) are only singular at the end-points A and B.


\subsection*{Semi-analytic solution with the double-exponential and the Gauss-Legendre quadratures}

\begin{figure}
\includegraphics[width=\columnwidth]{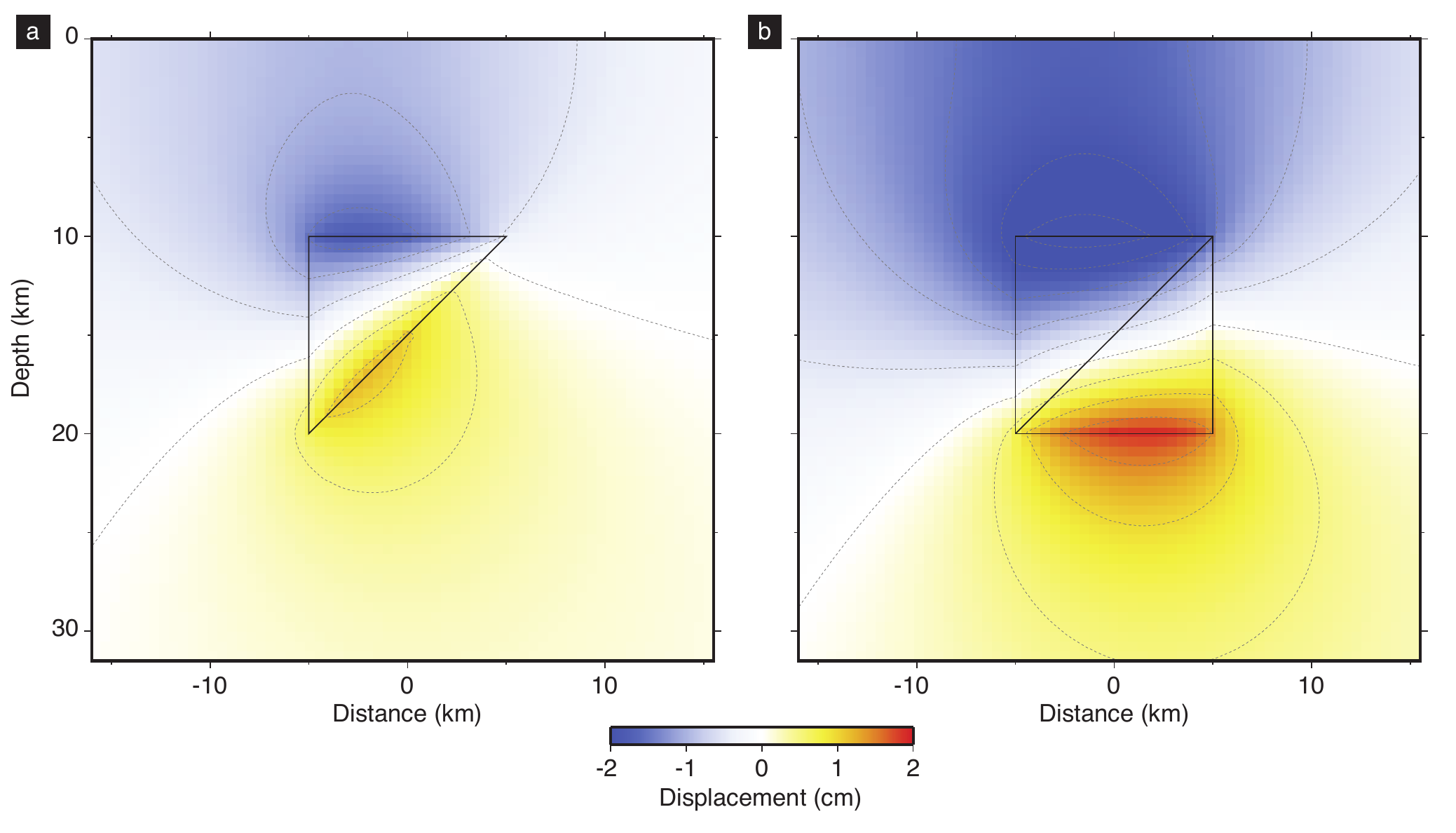}
\caption{Displacement field in anti-plane strain due to anelastic strain confined in triangular elements (triangles). A) A single triangular element and B) two triangle elements forming a rectangle. The strain volumes are subjected to the transformation strain $\epsilon_{12}=10^{-6}$ and $\epsilon_{13}=4\times 10^{-6}$. The contours (dashed lines) are every 5\,mm.}\label{fig:antiplane-triangle-rectangle}
\end{figure}

I obtain the solution semi-analytically by solving the line integrals using high-precision numerical quadratures. The Gauss-Legendre quadrature~\citep{golub+welsch69,abramowitz+stegun72} provides accurate solutions away from singular points. The double-exponential quadrature~\citep{haber77} is more robust to the presence of singularities. To proceed, I consider the line integral
\begin{equation}
I(x_2,x_3)=\int_{AB}G_{11}(x_2,x_3;y_2,y_3)\,\diff y_2\diff y_3~,
\end{equation}
which I write as a parameterized line integral and in the canonical form of the double-exponential or the Gauss-Legendre quadrature, i.e., within the bounds of integration $-1$ and $1$, to get
\begin{equation}\label{eqn:path-integral}
I(x_2,x_3)=\frac{R}{2}\int_{-1}^1G_{11}(x_2,x_3;y_2(t),y_3(t))\,\diff t~,
\end{equation}
where $R$ is the length of segment AB, $t$ is a dummy variable of integration, and
\begin{equation}
\begin{aligned}
y_2(t)&=\frac{x_2^A+x_2^B}{2}+t\,\frac{x_2^B-x_2^A}{2}~, \\
y_3(t)&=\frac{x_3^A+x_3^B}{2}+t\,\frac{x_3^B-x_3^A}{2}~. \\
\end{aligned}
\end{equation}
The displacement field for a combination of horizontal and vertical shear strain is shown in Figure~\ref{fig:antiplane-triangle-rectangle}. The numerical solution with the double-exponential quadrature agrees with the analytic solution (\ref{eqn:closed-form-antiplane}) within double-precision floating-point accuracy (about twelve digits) but takes about 50 times longer to evaluate. The Gauss-Legendre quadrature with just 15 integration points provides high precision in the far-field and can be evaluated almost as fast as the analytic solution. Therefore, switching from the double-exponential to the Gauss-Legendre method when the distance from the circumcenter exceeds 1.75 times the circumradius provides optimal performance without sacrificing accuracy. Two triangles can be combined to form a rectangle. In this case the analytic and semi-analytic solutions agree with the closed-form solution of~\cite{barbot+17}. 

\subsection*{Stress and strain}

\begin{figure}
\includegraphics[width=\columnwidth]{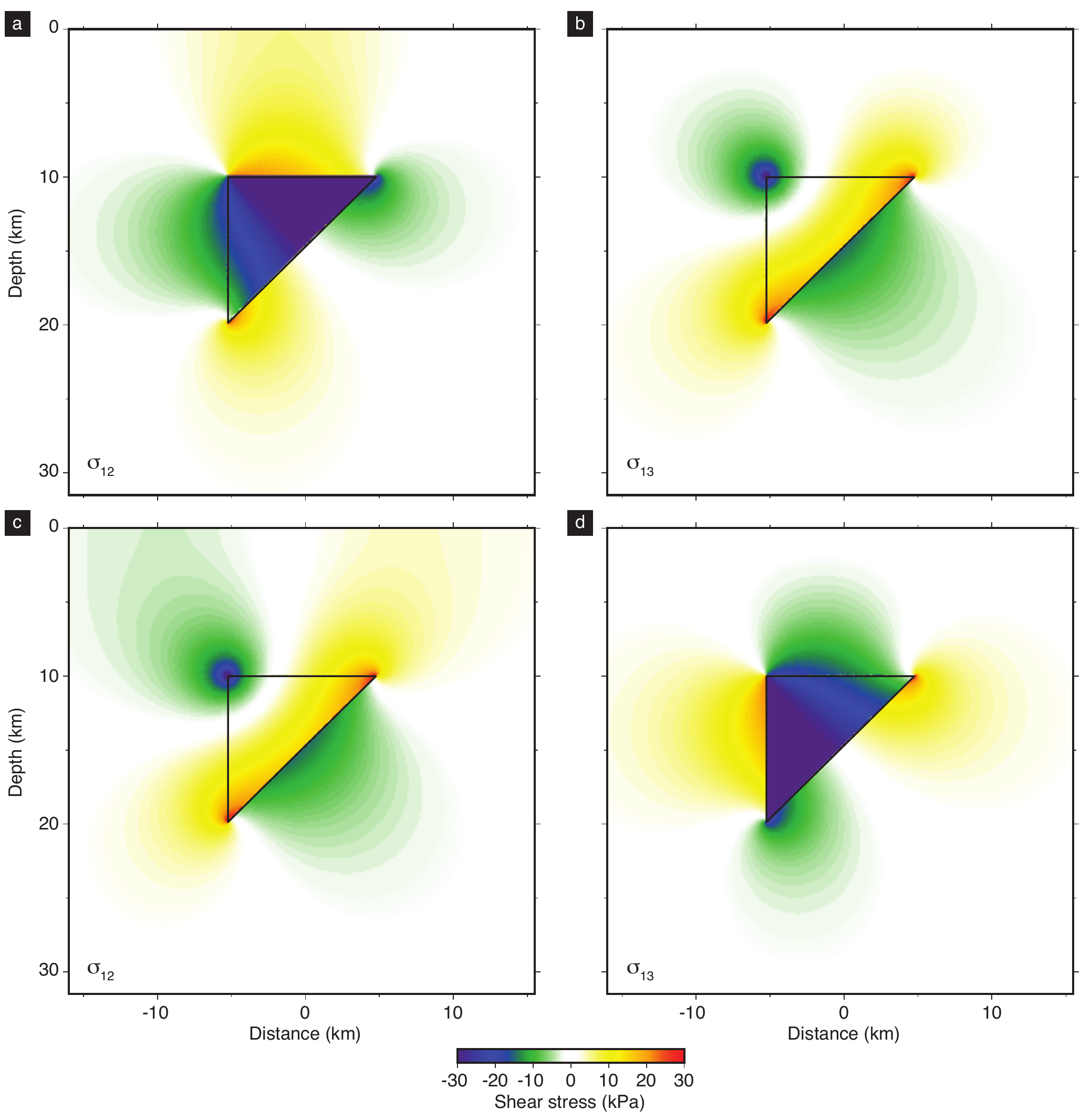}
\caption{Stress field in anti-plane strain due to anelastic strain confined in triangular elements (triangles). a, c) Horizontal shear stress $\sigma_{12}$ and b, d) Vertical shear stress $\sigma_{13}$. The strain volumes in the left panel are subjected to the transformation strain $\epsilon_{12}=10^{-6}$. In the right panel, to $\epsilon_{13}=10^{-6}$.}\label{fig:antiplane-triangle-stress}
\end{figure}

The stress field can be obtained using (\ref{eqn:stress-eigenstrain}). For a triangular region with vertices A, B, and C as in Figure~\ref{fig:geometry-antiplane-triangle}, the location of the transformation strain is given by
\begin{equation}\label{eqn:omega-triangle}
\begin{aligned}
\Phi(\textbf{x})=
&H\left[\left(\frac{\textbf{x}^A+\textbf{x}^B}{2}-\textbf{x}\right)\cdot\textbf{n}^{(C)}\right] \\
\times&H\left[\left(\frac{\textbf{x}^B+\textbf{x}^C}{2}-\textbf{x}\right)\cdot\textbf{n}^{(A)}\right] \\
\times&H\left[\left(\frac{\textbf{x}^C+\textbf{x}^A}{2}-\textbf{x}\right)\cdot\textbf{n}^{(B)}\right]~,
\end{aligned}
\end{equation}
where $H(x)$ is the Heaviside function. An example of the spatial distribution of the shear stress around a triangular strain volume in shown in Figure~\ref{fig:antiplane-triangle-stress}. When two triangles are combined to form a rectangle, it creates the stress field derived by \cite{barbot+17}. The semi-analytic solution agrees with the analytic expression based on (\ref{eqn:anti-plane-strain-analytic}) to double-precision floating point accuracy and takes a similar time to evaluate. This indicates that combining the double-exponential and the Gauss-Legendre quadratures is a viable approach when a closed-form solution is otherwise unavailable.

\section*{Distributed deformation of triangular strain regions in plane strain}

The dynamics of the lithosphere-asthenosphere system around subduction zones, normal faults, and spreading centers may be investigated under the plane strain approximation~\citep{sato+matsuura74,cohen96,savage98,hirahara02,liu+rice05,muto+13,dinther+13,govers+17,biemiller+lavier17,romanet+18,barbot18b,goswami+barbot18}. In particular, \cite{glas91} derived the closed expression for the displacement and stress due to a cuboidal inclusion aligned with the free surface and~\cite{barbot+17} expanded the results for a rotated cuboidal source. In this section, I develop closed-form analytic and semi-analytical solutions for triangular elements of arbitrary orientation to conform with curvilinear meshes. This type of element may prove useful to capture the geometry of the mantle wedge corner, of shear zones below volcanic arcs, or weak regions surrounding dykes and sills. 

\subsection*{Problem statement}
I consider the elastic deformation in plane strain caused by distributed anelastic strain in an elementary triangular area (Figure~\ref{fig:geometry-antiplane-triangle}). In plane strain, we have $u_{1,i}=0$ for $i=1,2,3$, and $u_1=0$. I consider a triangular area delineated by the vertices $A$, $B$, and $C$ and subjected to the transformation strain components $\epsilon^i_{22}$, $\epsilon^i_{23}$, $\epsilon^i_{32}$ and $\epsilon^i_{33}$, with $\epsilon_{23}^i=\epsilon_{32}^i$. Using (\ref{eqn:kernel-simplified}) again, the deformation simplifies to the nontrivial components
\begin{equation}
\begin{aligned}
u_2(x_2,x_3)=\int_{\partial\Omega}&G_{22}(x_2,x_3,y_2,y_3)\left(m_{22}n_2+m_{23}n_3\right) \\
&+G_{32}(x_2,x_3,y_2,y_3)\left(m_{32}n_2+m_{33}n_3\right)\,\diff y_2\diff y_3 \\
u_3(x_2,x_3)=\int_{\partial\Omega}&G_{23}(x_2,x_3,y_2,y_3)\left(m_{22}n_2+m_{23}n_3\right) \\
&+G_{33}(x_2,x_3,y_2,y_3)\left(m_{23}n_2+m_{33}n_3\right)\,\diff y_2\diff y_3~, \\
\end{aligned}
\end{equation}
where $G_{22}$ and $G_{23}$ represent the displacements at $(x_2,x_3)$ induced by a line force in the $\textbf{e}_2$ direction centered at $(y_2,y_3)$ and $G_{32}$ and $G_{33}$ represent the displacements induced by a line force in the $\textbf{e}_3$ direction. They are given by~\citep{melan32,dundurs62,segall10}
\begin{equation}\label{eqn:greens-function-2d}
\begin{aligned}
G_{22}=\frac{-1}{2\pi\,\mu(1-\nu)}&\bigg[\frac{3-4\nu}{4}\ln r_1 + \frac{8\nu^2-12\nu+5}{4}\ln r_2+\frac{(x_3-y_3)^2}{4\,{r_1}^2}\\
&\quad+\frac{(3-4\nu)(x_3+y_3)^2+2y_3\,(x_3+y_3)-2y_3^2}{4\,{r_2}^2}-\frac{y_3x_3\,(x_3+y_3)^2}{{r_2}^4}\bigg]\\
G_{23}=\frac{1}{2\pi\,\mu(1-\nu)}&\bigg[(1-2\nu)(1-\nu)\tan^{-1}\frac{x_2-y_2}{x_3+y_3}+\frac{(x_3-y_3)\,(x_2-y_2)}{4\,{r_1}^2}\\
&\quad+(3-4\nu)\frac{(x_3-y_3)\,(x_2-y_2)}{4\,{r_2}^2}-\frac{y_3x_3(x_2-y_2)\,(x_3+y_3)}{{r_2}^4}\bigg]~, \\
G_{32}=\frac{1}{2\pi\,\mu(1-\nu)}&\bigg[-(1-2\nu)(1-\nu)\tan^{-1}\frac{x_2-y_2}{x_3+y_3}+\frac{(x_3-y_3)\,(x_2-y_2)}{4\,{r_1}^2}\\
&\quad+(3-4\nu)\frac{(x_3-y_3)\,(x_2-y_2)}{4\,{r_2}^2}+\frac{y_3x_3(x_2-y_2)\,(x_3+y_3)}{{r_2}^4}\bigg]\\
G_{33}=\frac{1}{2\pi\,\mu(1-\nu)}&\bigg[-\frac{3-4\nu}{4}\ln r_1 - \frac{8\nu^2-12\nu+5}{4}\ln r_2\\
&\quad-\frac{(x_2-y_2)^2}{4\,{r_1}^2}+\frac{2y_3x_3-(3-4\nu)(x_2-y_2)^2}{4\,{r_2}^2}-\frac{y_3x_3\,(x_2-y_2)^2}{{r_2}^4}\bigg]~,\\
\end{aligned}
\end{equation}
with the radii
\begin{equation}
\begin{aligned}
r_1^2&=(x_2-y_2)^2+(x_3-y_3)^2\\
r_2^2&=(x_2-y_2)^2+(x_3+y_3)^2~.
\end{aligned}
\end{equation}
Breaking down the path integral along the three triangle segments, it becomes
\begin{equation}
\begin{aligned}
u_2(x_2,x_3)
&=\left(m_{22}n_2^{(C)}+m_{23}n_3^{(C)}\right)\int_{AB}G_{22}(x_2,x_3;y_2,y_3)\,\diff y_2\diff y_3 \\
&+\left(m_{32}n_2^{(C)}+m_{33}n_3^{(C)}\right)\int_{AB}G_{32}(x_2,x_3;y_2,y_3)\,\diff y_2\diff y_3 \\
&+\left(m_{22}n_2^{(A)}+m_{23}n_3^{(A)}\right)\int_{BC}G_{22}(x_2,x_3;y_2,y_3)\,\diff y_2\diff y_3 \\
&+\left(m_{32}n_2^{(A)}+m_{33}n_3^{(A)}\right)\int_{BC}G_{32}(x_2,x_3;y_2,y_3)\,\diff y_2\diff y_3 \\
&+\left(m_{22}n_2^{(B)}+m_{23}n_3^{(B)}\right)\int_{AC}G_{22}(x_2,x_3;y_2,y_3)\,\diff y_2\diff y_3 \\
&+\left(m_{32}n_2^{(B)}+m_{33}n_3^{(B)}\right)\int_{AC}G_{32}(x_2,x_3;y_2,y_3)\,\diff y_2\diff y_3~, \\
\end{aligned}
\end{equation}
and
\begin{equation}
\begin{aligned}
u_3(x_2,x_3)
&=\left(m_{22}n_2^{(C)}+m_{23}n_3^{(C)}\right)\int_{AB}G_{23}(x_2,x_3;y_2,y_3)\,\diff y_2\diff y_3 \\
&+\left(m_{32}n_2^{(C)}+m_{33}n_3^{(C)}\right)\int_{AB}G_{33}(x_2,x_3;y_2,y_3)\,\diff y_2\diff y_3 \\
&+\left(m_{22}n_2^{(A)}+m_{23}n_3^{(A)}\right)\int_{BC}G_{23}(x_2,x_3;y_2,y_3)\,\diff y_2\diff y_3 \\
&+\left(m_{32}n_2^{(A)}+m_{33}n_3^{(A)}\right)\int_{BC}G_{33}(x_2,x_3;y_2,y_3)\,\diff y_2\diff y_3 \\
&+\left(m_{22}n_2^{(B)}+m_{23}n_3^{(B)}\right)\int_{AC}G_{23}(x_2,x_3;y_2,y_3)\,\diff y_2\diff y_3 \\
&+\left(m_{32}n_2^{(B)}+m_{33}n_3^{(B)}\right)\int_{AC}G_{33}(x_2,x_3;y_2,y_3)\,\diff y_2\diff y_3~, \\
\end{aligned}
\end{equation}
where $\textbf{n}^{(A)}$, $\textbf{n}^{(B)}$, and $\textbf{n}^{(C)}$ are the unit normal vectors to the sides BC, AC, and AB, respectively.

\subsection*{Analytic and semi-analytic solutions}

\begin{figure}[p]
\includegraphics[width=\columnwidth]{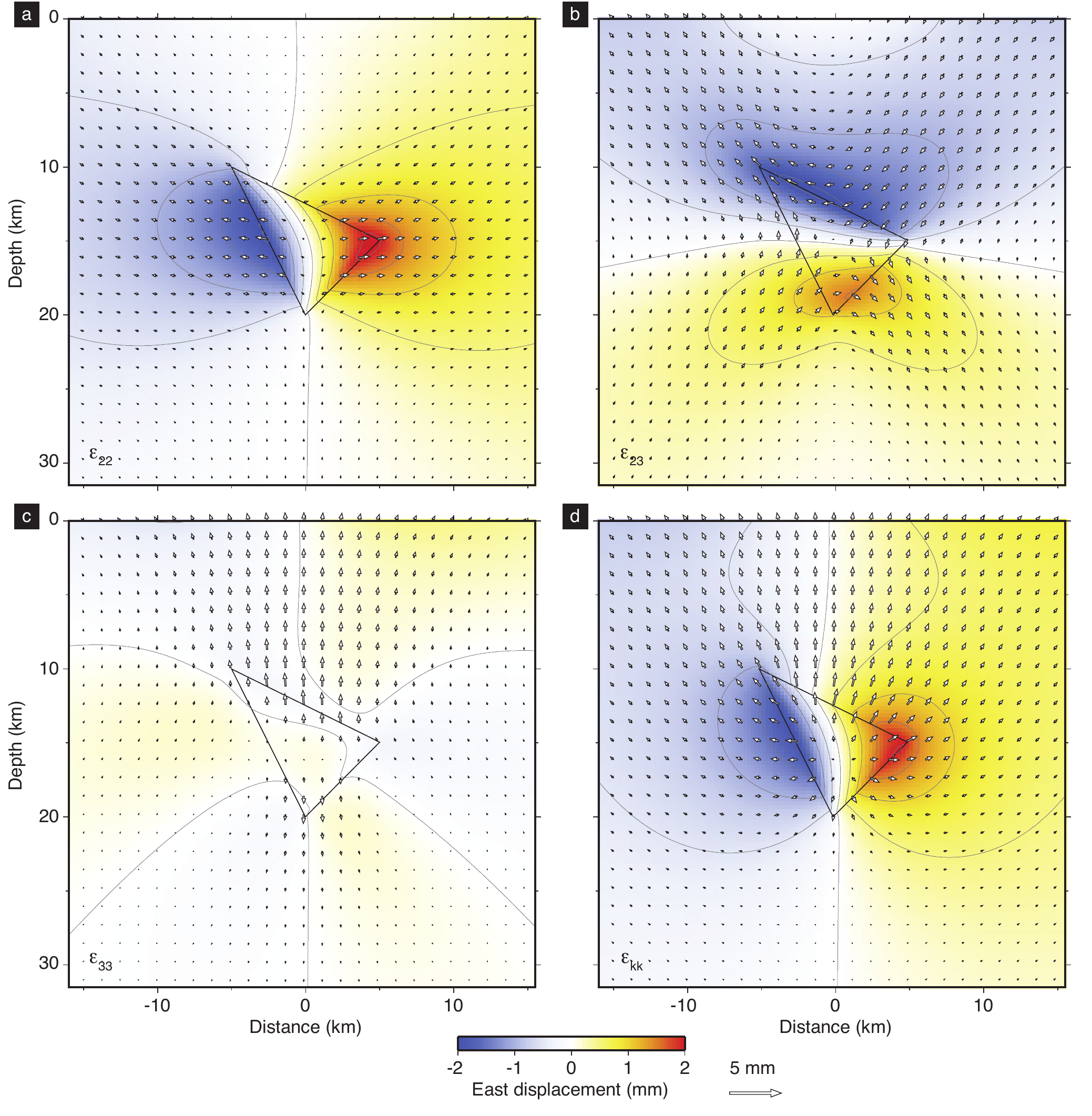}
\caption{Displacement field in plane strain strain due to anelastic strain confined in triangular elements (triangles). The horizontal displacement is shown by the arrows and the color indicates the horizontal component $u_2$. The displacement fields are due to a) horizontal uniaxial extension ($\epsilon_{22}=10^{-6}$), b) pure shear ($\epsilon_{23}=10^{-6}$), c) vertical uniaxial extension ($\epsilon_{33}=10^{-6}$), and d) isotropic extension ($\epsilon_{22}=\epsilon_{33}=10^{-6}$). The contours (dashed lines) are every 0.5\,mm.}\label{fig:planestrain-triangle-eij}
\end{figure}

\begin{figure}[p]
\includegraphics[width=\columnwidth]{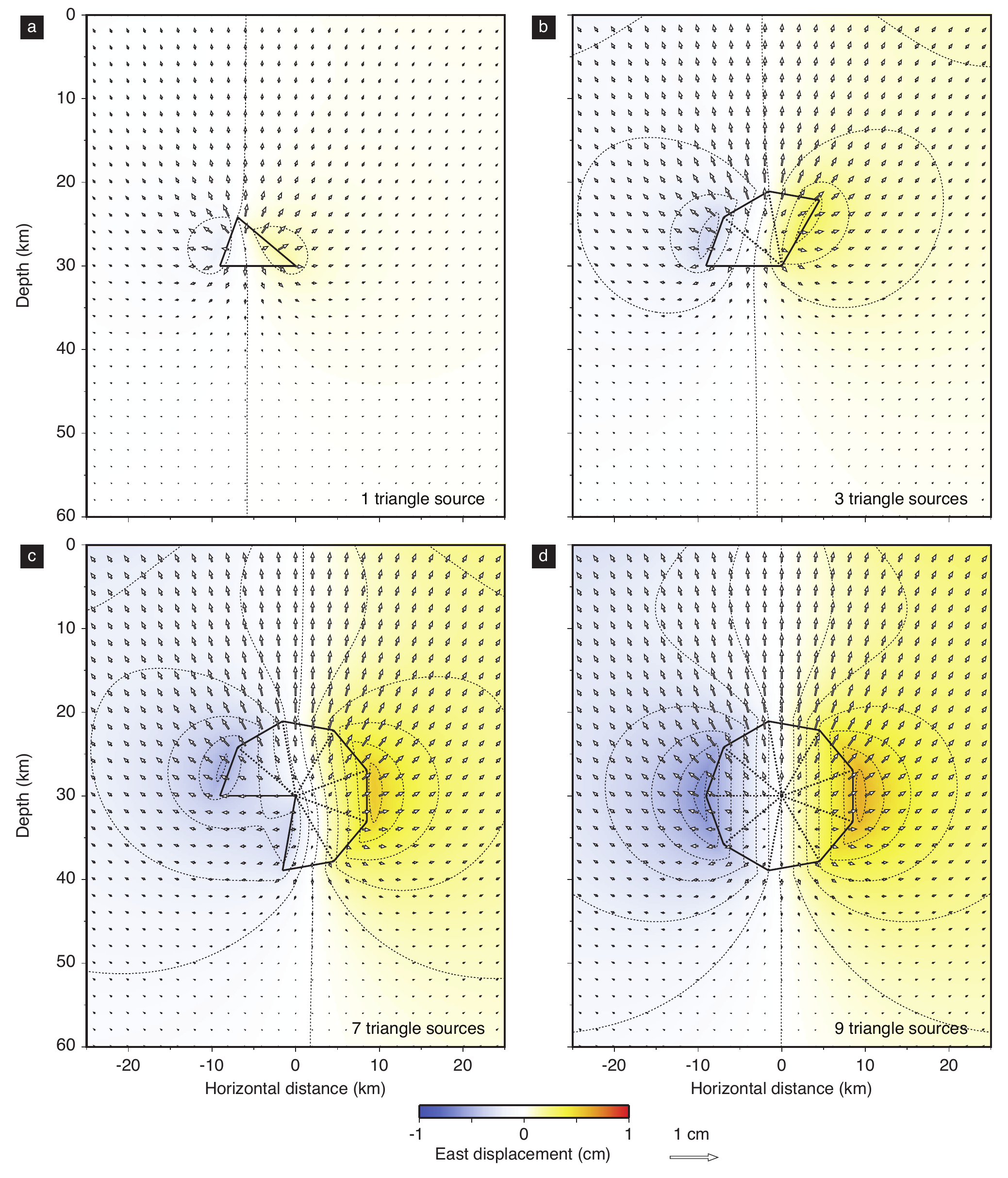}
\caption{Approximation of the displacement field for a dilating disk by combining 9 triangular sources. The triangles are form by joining points around the surrounding circle with the center, here at 30\,km depth. The background shows the amplitude of the horizontal displacement. The line integrals of the shared segments of adjacent triangles (dashed segments) cancel out, leading to the outwards segments of the combined volume having a non-trivial contribution. a) Displacement field due to an elementary triangle of dilatation. b) Cumulative displacement field due to three triangles. c) Case for 7 triangles. d) Displacement field approximated with 9 triangular sources. The dashed contours are every 1\,mm of horizontal displacement.}\label{fig:planestrain-circle-triangle-short}
\end{figure}

The displacement field can be evaluated analytically or using a numerical quadrature using the path integral of the form (\ref{eqn:path-integral}). The closed-form expressions for the line integrals
\begin{equation}\label{eqn:closed-form-in-plane-strain}
U_{ij}=\int_{AB}G_{ij}(x_2,x_3;y_2,y_3)\,\diff y_2\diff y_3~,
\end{equation} 
for the plane-strain Green's functions (\ref{eqn:greens-function-2d}) and $i=2,3$ are provided in Appendix A. Examples of displacement fields occasioned by distributed anelastic strain confined in the triangle surface ABC are given in Figure~\ref{fig:planestrain-triangle-eij}. In Figure~\ref{fig:planestrain-circle-triangle-short}, I show how triangles can be combined to approximate a disk in dilatation. I have checked that combining two triangles to form a rectangle conforms to the analytic solution of~\cite{barbot+17} and that the numerical integration with the double-exponential and the Gauss-Legendre quadratures agrees with the closed-form solution of (\ref{eqn:closed-form-in-plane-strain}) up to double-precision floating point accuracy for all combinations of sources and displacement components.

\subsection*{Stress and strain}

The strain can be obtained by differencing the displacement field analytically or with a finite-difference approximation. An alternative is to directly integrate the Green's functions for the displacement gradient, given below

\begin{equation}\label{eqn:greens2dderivatives1}
\begin{aligned}
G_{22,2}&=-\frac{x_2-y_2}{2\pi G(1-\nu)}\bigg[
\frac{3-4\nu}{4\,{r_1}^2}  +\frac{8\nu^2-12\nu+5}{4\,{r_2}^2} -\frac{(x_3-y_3)^2}{2\,{r_1}^4} \\
&\quad\qquad\qquad\qquad -\frac{(3-4\nu)(x_3+y_3)^2+2y_3(x_3+y_3)-2{y_3}^2}{2\,{r_2}^4} \\
&\quad\qquad\qquad\qquad  + \frac{4\,y_3x_3(x_3+y_3)^2}{{r_2}^6} \bigg]  \\
G_{22,3}&=-\frac{1}{2\pi G(1-\nu)} \bigg[ {\frac { \left( 3-4\,\nu \right)  \left( x_3- y_3 \right) }{{{ 4\,r_1}}^{2}}}+{\frac { \left( 8\,{\nu}^{2}-12\,\nu+5 \right)  \left( 
{ x_3}+{ y_3} \right) }{{{ 4\,r_2}}^{2}}} \\
&\quad\qquad\qquad\qquad+{\frac {{ x_3}-{ y_3}}{{{ 2\,r_1}}^{2}}}-{\frac { \left( { x_3}-{ y_3} \right) ^{3}  }{{{ 2\,r_1}}^{4}}} +{\frac {\left( 3-4\,\nu \right)  \left( { x_3}+{ y_3} \right) +{ y_3}}{{{ 2\,r_2}}^{2}}} \\
 &\quad\qquad\qquad\qquad-\left( { x_3}+{ y_3
} \right){\frac {   \left( 3-4\,\nu \right)  \left( { x_3}+{ y_3} \right) ^{2}+2\, { y_3}  \left( { x_3}+{ y_3} \right) -2\,{{ y_3}}^{2}  }{{{ 2\,r_2}}^{4}}} \\
&\quad\qquad\qquad\qquad-{\frac {  { y_3} \left( { x_3}+{ y_3} \right) ^{2}}{{{ r_2}}^{4}}} -2\,{\frac {  { y_3}  \,{ x_3} \left( { x_3}+{ y_3} \right) }{{{ r_2}}^{4}}} \\
&\quad\qquad\qquad\qquad+4\,{\frac {{ x_3}\,   y_3 \left( { x_3}+{ y_3} \right) ^{3}  }{{{ r_2}}^{6}}} \bigg] \\
\end{aligned}
\end{equation}
\begin{equation}\label{eqn:greens2dderivatives2}
\begin{aligned}
G_{23,2}&=\frac{1}{2\pi G (1-\nu)} \bigg[  \left( 1-2\,\nu \right)  \left( 1-\nu \right)  \frac{ {x_3}+{y_3} }{{r_2}^2} +{\frac {{x_3}-{y_3}}{4\,{r_1}^{2}}} \\
&\quad\qquad\qquad\qquad-{\frac { \left( {x_3}-{y_3} \right)  \left( {x_2}-{y_2} \right)^2 }{2\,{r_1}^{4}}}+\left( 3-4\,\nu \right){\frac {    {x_3}-{y_3}  }{4\,{r_2}^{2}}} \\
&\quad\qquad\qquad\qquad-{\left( 3-4\,\nu \right)\frac {   \left( {x_3}-{y_3} \right)  \left( {x_2}-{y_2} \right)^2   }{2\,{r_2}^{4}}}-{\frac {  y_3 \,{x_3}\, \left( {x_3}+{y_3} \right) }{{{r_2}}^{4}}} \\
&\quad\qquad\qquad\qquad+4\,{\frac { y_3  \,{x_3}\, \left( {x_2}-{y_2} \right)^2  \left( {x_3}+{y_3} \right)   }{{{r_2}}^{6}}} \bigg]~, \\
G_{23,3}&=\frac{{ x_2}-{ y_2}}{2\pi G (1-\nu)} \bigg[ -\left( 1-2\,\nu \right)  \left( 1-\nu \right)  \frac{ 1} {{r_2}^2}+{\frac {1}{4\,{r_1}^{2}}} \\
&\quad\qquad\qquad\qquad-{\frac { \left( { x_3}-{ y_3} \right)^2 }{2\,{r_1}^{4}}}+{\frac { \left( 3-4\,\nu \right)  }{4\,{
r_2}^{2}}} \\
&\quad\qquad\qquad\qquad-{\frac { \left( 3-4\,\nu \right)  \left( { x_3}-{ y_3} \right)  \left( x_3+y_3 \right) }{2\,{r_2}^{4}}
} \\
&\quad\qquad\qquad\qquad-{\frac { y_3 \left( { x_3}+{ y_3} \right) }{{{ r_2}}^{4}}}-{\frac { y_3 \,{ x_3}\,  }{{{ r_2}}^{4}}} \\
 &\quad\qquad\qquad\qquad+4\,{\frac { y_3 \,{ x_3}\,   \left( { x_3}+{ y_3} \right)^2 }{{{ r_2}}^{6}}} \bigg]~,  \\
 \end{aligned}
 \end{equation}
\begin{equation}\label{eqn:greens2dderivatives3}
\begin{aligned}
G_{32,2}&= \frac{1}{2\pi G (1-\nu)} \bigg[ - \left( 1-2\,\nu \right)  \left( 1-\nu \right)  \frac{ x_3+ y_3}{{r_2}^2} +{\frac {{ x_3}-{ y_3}}{4\,{r_1}^{2}}} \\
&\quad\qquad\qquad\qquad-{\frac { \left( { x_3}-{ y_3} \right)  \left( { x_2}-{ y_2} \right)^2  }{2\,{r_1}^{4}}}+{\frac { \left( 3-4\,\nu \right)  \left( { x_3}-{ y_3} \right) }{4\,{r_2}^{2}}} \\
&\quad\qquad\qquad\qquad-{\frac { \left( 3-4\,\nu \right)  \left( { x_3}-{ y_3} \right)  \left( { x_2}-{ y_2} \right)^2 }{2\,{r_2}^{4}}}+{\frac { y_3\,{ x_3}\left( { x_3}+{ y_3} \right) }{{{ r_2}}^{4}}} \\
&\quad\qquad\qquad\qquad-4\,{\frac { y_3\,{ x_3}\left( { x_2}-{ y_2} \right)^2 \left( { x_3}+{ y_3} \right)   }{{{ r_2}}^{6}}} \bigg]~, \\
G_{32,3}&= \frac{x_2- y_2}{2\pi G (1-\nu)} \bigg[  \left( 1-2\,\nu \right)  \left( 1-\nu \right)  \frac{ 1}{{r_2}^2} +{\frac {1}{4\,{r_1}^{2}}} \\
 &\quad\qquad\qquad\qquad-{\frac { \left( { x_3}-{ y_3} \right)^2  }{2\,{r_1}^{4}}}+\left( 3-4\,\nu \right){\frac { 1   }{4\,{r_2}^{2}}} \\
 &\quad\qquad\qquad\qquad-\left( 3-4\,\nu \right){\frac {   \left( { x_3}-{ y_3} \right)    \left( x_3+y_3 \right) }{2\,{r_2}^{4}}} \\
 &\quad\qquad\qquad\qquad+{\frac { y_3 \left( { x_3}+{ y_3} \right) }{{{ r_2}}^{4}}}+{\frac { y_3\,{ x_3} }{{{ r_2}}^{4}}} \\
 &\quad\qquad\qquad\qquad-4\,{\frac { y_3 \,{ x_3}  \left( { x_3}+{ y_3} \right)^2  }{{{ r_2}}^{6}}} \bigg]~, \\
 \end{aligned}
 \end{equation}
\begin{equation}\label{eqn:greens2dderivatives4}
\begin{aligned}
G_{33,2}&= -\frac{ x_2- y_2 }{2\pi G (1-\nu)} \bigg[  \left( 3-4\,\nu \right){\frac {   1 }{4\,{r_1}^{2}}}+\left( 8\,{\nu}^{2}-12\,\nu+5 \right){\frac {   1}{4\,{r_2}^{2}}} \\
&\quad\qquad\qquad\qquad+{\frac {1}{2\,{r_1}^{2}}}-{\frac { \left( { x_2}-{ y_2} \right) ^{2}  }{2\,{r_1}^{4}}} \\
&\quad\qquad\qquad\qquad+\left( 3-4\,\nu \right){\frac {   1 }{2\,{r_2}^{2}}} \\
&\quad\qquad\qquad\qquad+{\frac { 2\,  y_2 \,{ x_3}- \left( 3-4\,\nu \right)  \left( { x_2}-{ y_2} \right) ^{2}  }{2\,{r_2}^{4}}} \\
&\quad\qquad\qquad\qquad+2\,{\frac {  y_3 \,{ x_3}  }{{{ r_2}}^{4}}}-4\,{\frac {{ x_3}\,  y_3  \left( { x_2}-{ y_2} \right) ^{2}  }{{{ r_2}}^{6}}} \bigg]~, \\
G_{33,3}&= \frac{1}{2\pi G (1-\nu)} \bigg[   -{\frac { \left( 3-4\,\nu \right)  \left( x_3-y_3 \right) }{4\,{r_1}^{2}}}-{\frac { \left( 8\,{\nu}^{2}-12\,\nu+5 \right)  \left( x_3+y_3 \right) }{4\,{r_2}^{2}}} \\
&\quad\qquad\qquad\qquad+{\frac { \left( { x_2}-{ y_2} \right) ^{2} \left( x_3-y_3\right) }{2\,{r_1}^{4}}}+{\frac {{ y_2}}{2\,{r_2}^{2}}} \\
 &\quad\qquad\qquad\qquad-\left( x_3+y_3\right){\frac { 2\, y_2\,{ x_3}- \left( 3-4\,\nu \right)  \left( { x_2}-{ y_2} \right) ^{2}}{2\,{r_2}^{4}}} \\
&\quad\qquad\qquad\qquad-{\frac { y_3\left( { x_2}-{ y_2} \right) ^{2}}{{{ r_2}}^{4}}}+4\,{\frac {{ x_3}\, y_3 \left( { x_2}-{ y_2} \right) ^{2} \left( x_3+y_3 \right) }{{{ r_2}}^{6}}} \bigg]~,
\end{aligned}
\end{equation}
where the comma in expressions like $G_{ij,k}$ indicates differentiation of the tensor component $G_{ij}$ with respect to $x_k$. Importantly, in all the terms forming the Green's function  (\ref{eqn:greens-function-2d}) and their derivatives (\ref{eqn:greens2dderivatives1}-\ref{eqn:greens2dderivatives4}), the only singular point is at $\textbf{x}=\textbf{y}$. This guarantees that the displacement and stress solutions based on numerical integration of the Green's function will be numerically stable at any point away from the source. This property provides an appealing reason to resort to numerical quadratures because, contrarily to many analytic solutions (including the one in Appendix~A), all points away from the contour of the source region are numerically stable.

\begin{figure}[p]
\includegraphics[width=\columnwidth]{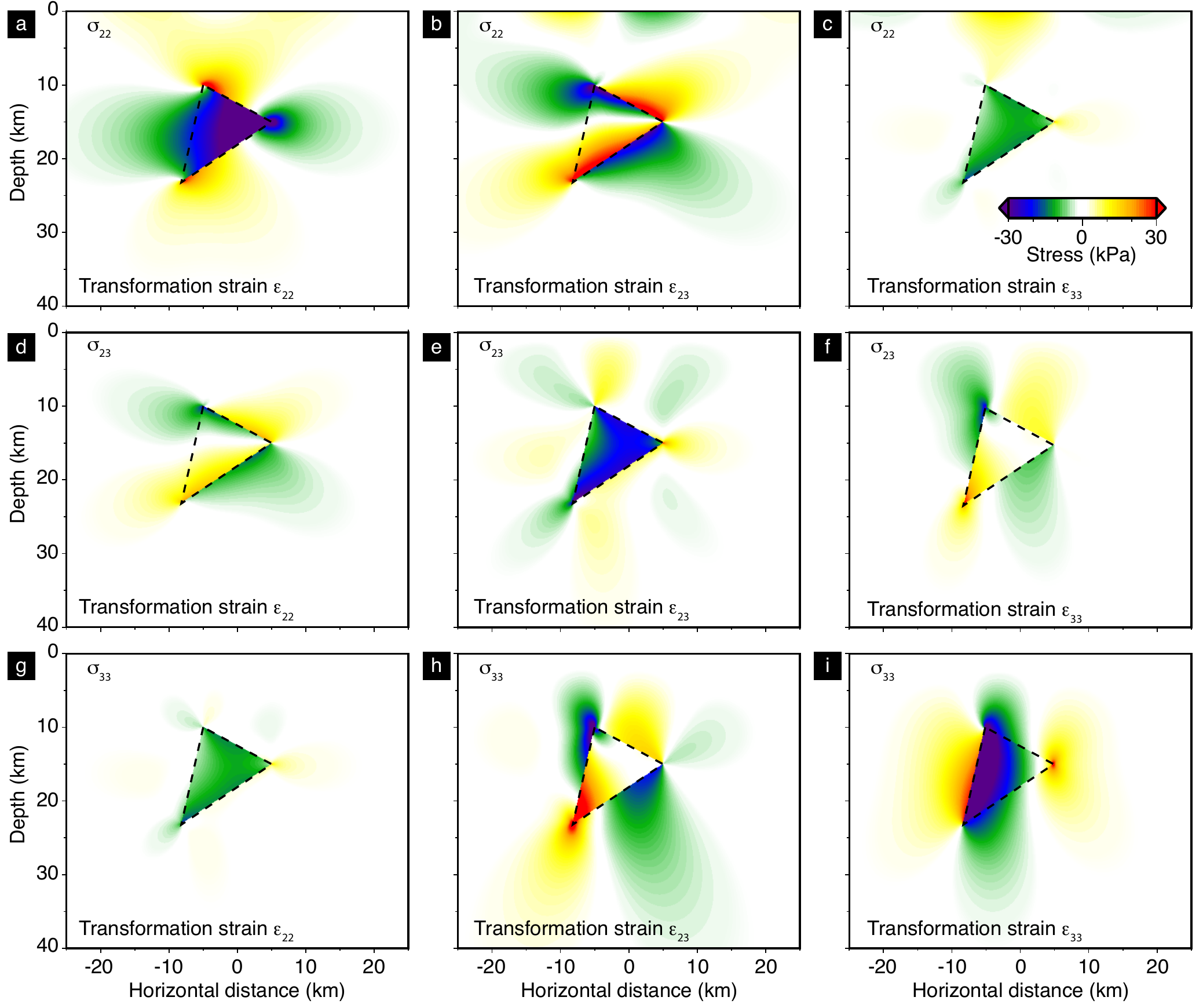}
\caption{Stress field in plane strain strain due to anelastic strain confined in triangular elements (dashed triangles). The top panel with a), b), and c) show the horizontal stress. The middle panel with d), e), and f) show the shear stress. The bottom panel with g), h), and i) show the vertical stress. The left column with a), d), and g) are for the horizontal transformation strain $\epsilon_{22}=10^{-6}$. The middle column with b), e), and h) are for a shear transformation strain of $\epsilon_{23}=10^{-6}$. The right column with c), f), and i) is for a vertical transformation strain of $\epsilon_{33}=10^{-6}$.}\label{fig:stress-planestrain-triangle}
\end{figure}

The $\Phi(\textbf{x})$ function that describes the location of transformation strain is the same for anti-plane and in-plane strain problems, so the stress and strain components can be obtained using (\ref{eqn:displacement-strain}), (\ref{eqn:stress-eigenstrain}), and (\ref{eqn:omega-triangle}). Figure~\ref{fig:stress-planestrain-triangle} shows the combination of the 3 stress components obtained by deformation of a triangular source with 3 different components of transformation strain. I have checked that the results from the finite-difference and the numerical quadrature methods converge for these 9 cases.

\clearpage
\section*{Distributed deformation of tetrahedral strain volumes in three dimensions}

The development of three-dimensional deformation models has afforded an increasingly accurate description of the mechanics of the lithosphere~\citep{mansinha+smylie71,sato+matsuura74,wang+03a,okada85,okada92,meade07a,nikkhoo+15,mctigue&segall88,aagaard+13,barbot+17,landry+barbot16}. The expressions for the deformation induced by uniform transformation strain confined in a cuboid have been developed for a full elastic medium by~\cite{faivre69}. \cite{chiu78} derived the solution for certain components of displacement and strain at the surface of a half-space. \cite{barbot+17} derived the displacement and stress everywhere in a half-space. The development of realistic rheological models of Earth's interior requires curvilinear meshes that conform to structural data, so in this manuscript I derive solutions for the case of transformation strain confined in a tetrahedral volume in a half-space. 

\subsection*{Problem statement}

\begin{figure}
\includegraphics[width=\columnwidth]{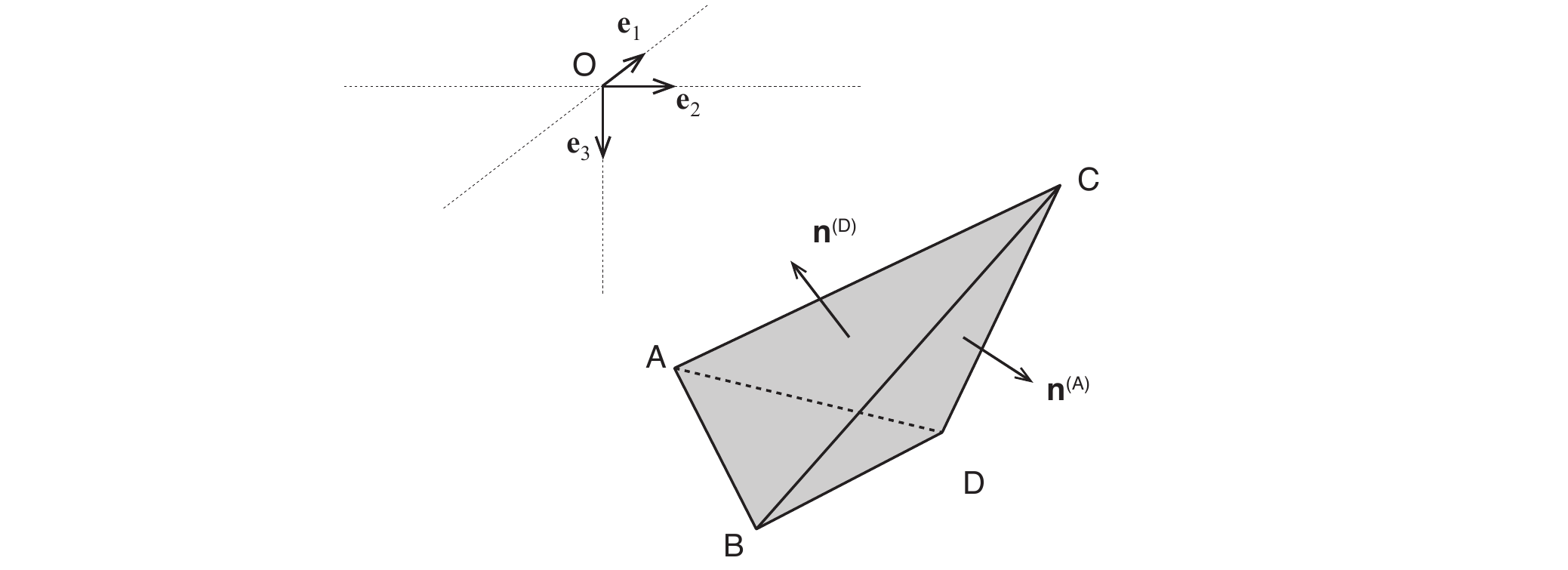}
\caption{Three-dimensional deformation of a half-space due to anelastic strain confined in a tetrahedral element with vertices A, B, C, and D at coordinates $\textbf{x}^A$, $\textbf{x}^B$, $\textbf{x}^C$, and $\textbf{x}^D$, respectively. The normal vectors are pointing outwards, such that $\textbf{n}^{(D)}\cdot(\textbf{x}_D-\textbf{x}_A)\le0$, $\textbf{n}^{(D)}\cdot(\textbf{x}_D-\textbf{x}_B)\le0$, and $\textbf{n}^{(D)}\cdot(\textbf{x}_D-\textbf{x}_C)\le0$.}\label{fig:geometry-3d-tetrahedron}
\end{figure}

I now consider deformation in a three-dimensional half-space (Figure~\ref{fig:geometry-3d-tetrahedron}). I consider a tetrahedral volume delineated by the vertices A, B, C, and D at coordinates $\textbf{x}^A$, $\textbf{x}^B$, $\textbf{x}^C$, and $\textbf{x}^D$, respectively, and subjected to the six independent transformation strain components
\begin{equation}
\boldsymbol{\epsilon}^i=\left(\begin{matrix}
\epsilon_{11} & \epsilon_{12} & \epsilon_{13} \\
\epsilon_{12} & \epsilon_{22} & \epsilon_{23} \\
\epsilon_{13} &  \epsilon_{23} &  \epsilon_{33} \\
\end{matrix}\right)~.
\end{equation}
Using (\ref{eqn:kernel-simplified}), the deformation simplifies to
\begin{equation}\label{eqn:surface-integral-3d}
\begin{aligned}
u_i(x_1,x_2,x_3)=
  \int_{\partial\Omega}&G_{1i}(x_1,x_2,x_3;y_1,y_2,y_3)\left(m_{11}n_1+m_{12}n_2+m_{13}n_3\right)\,\diff y_1\diff y_2\diff y_3 \\
+\int_{\partial\Omega}&G_{2i}(x_1,x_2,x_3;y_1,y_2,y_3)\left(m_{12}n_1+m_{22}n_2+m_{23}n_3\right)\,\diff y_1\diff y_2\diff y_3 \\
+\int_{\partial\Omega}&G_{3i}(x_1,x_2,x_3;y_1,y_2,y_3)\left(m_{13}n_1+m_{32}n_2+m_{33}n_3\right)\,\diff y_1\diff y_2\diff y_3
\end{aligned}
\end{equation}
for $i=1,2,3$, where $G_{ij}(\textbf{x};\textbf{y})$ represents the displacement component $u_j(\textbf{x})$ induced by a point force in the $\textbf{e}_i$ direction located at $\textbf{y}$. The Green's functions for the $u_1$ component are given by~\citep{mindlin36,press65,okada85,segall10}
\begin{equation}\label{eqn:greens3d1}
\begin{aligned}
G_{11}&=\frac{1}{16\pi\mu(1-\nu)}\bigg[\frac{3-4\nu}{R_1}+\frac{1}{R_2}+\frac{{(x_1-y_1)}^2}{{R_1}^3} \\
&\qquad\qquad\qquad\qquad+\frac{(3-4\nu)(x_1-y_1)^2}{{R_2}^3} \\
&\qquad\qquad\qquad\qquad+\frac{2\,x_3y_3\left({R_2}^2-3\,(x_1-y_1)^2\right)}{{R_2}^5}\\
&\qquad\qquad\qquad\qquad+\frac{4 (1-2\nu)(1-\nu)\left({R_2}^2-(x_1-y_1)^2+R_2 \left(x_3+y_3\right)\right)}{R_2 \left(R_2+x_3+y_3\right)^2} \bigg]~,\\
G_{21}&=\frac{\left(x_1-y_1\right) \left(x_2-y_2\right)}{16\pi\mu(1-\nu )}\bigg[\frac{1}{{R_1}^3}+\frac{3-4\nu}{{R_2}^3}-\frac{6\, x_3 y_3}{{R_2}^5}-\frac{4 (1-2\nu) (1-\nu)}{R_2 \left(R_2+x_3+y_3\right)^2}\bigg]~,\\
G_{31}&=\frac{\left(x_1-y_1\right)}{16\pi\mu(1-\nu)} \bigg[\frac{x_3-y_3}{{R_1}^3}+\frac{(3-4\nu)\left(x_3-y_3\right)}{{R_2}^3} \\
&\qquad\qquad\qquad\qquad+\frac{6\, x_3 y_3 \left(x_3+y_3\right)}{{R_2}^5}-\frac{4\,(1-2\nu) (1-\nu)}{R_2\left(R_2+x_3+y_3\right)}\bigg]~.\\
\end{aligned}
\end{equation}
For the $u_2$ component, they are
\begin{equation}
\begin{aligned}
G_{12}&=\frac{\left(x_1-y_1\right) \left(x_2-y_2\right)}{16 \pi  \mu  (1-\nu)} \bigg[\frac{1}{{R_1}^3}+\frac{3-4\nu}{{R_2}^3}-\frac{6\, x_3 y_3}{{R_2}^5}-\frac{4 (1-2\nu) (1-\nu)}{R_2 \left(R_2+x_3+y_3\right)^2}\bigg]~, \\
G_{22}&=\frac{1}{16\pi\mu (1-\nu)}\bigg[\frac{3-4 \nu }{R_1}+\frac{1}{R_2} \\
&\qquad\qquad\qquad\qquad+\frac{\left(x_2-y_2\right)^2}{{R_1}^3}+\frac{(3-4 \nu ) \left(x_2-y_2\right)^2}{{R_2}^3} \\
&\qquad\qquad\qquad\qquad+\frac{2\,x_3 y_3 \left({R_2}^2-3 \left(x_2-y_2\right)^2\right)}{{R_2}^5}\\
&\qquad\qquad\qquad\qquad+\frac{4 (1-2 \nu ) (1-\nu ) \left({R_2}^2-\left(x_2-y_2\right)^2+R_2\left(x_3+y_3\right)\right)}{R_2\left(R_2+x_3+y_3\right)^2}\bigg]~,\\
G_{32}&=\frac{\left(x_2-y_2\right)} {16\pi\mu(1-\nu)}\bigg[\frac{x_3-y_3}{{R_1}^3}+\frac{(3-4 \nu ) \left(x_3-y_3\right)}{{R_2}^3} \\
&\qquad\qquad\qquad\qquad+\frac{6\, x_3 y_3 \left(x_3+y_3\right)}{{R_2}^5}-\frac{4\,(1-2\nu)(1-\nu )}{R_2 \left(R_2+x_3+y_3\right)}\bigg]~.\\
\end{aligned}
\end{equation}
For the displacement component $u_3$, they are given by
\begin{equation}\label{eqn:greens3d3}
\begin{aligned}
G_{13}&=\frac{\left(x_1-y_1\right)}{16\pi\mu(1-\nu)}\bigg[\frac{x_3-y_3}{{R_1}^3}+\frac{(3-4 \nu ) \left(x_3-y_3\right)}{{R_2}^3} \\
&\qquad\qquad\qquad\qquad-\frac{6\, x_3 y_3 \left(x_3+y_3\right)}{{R_2}^5}+\frac{4(1-2\nu)(1-\nu)}{R_2 \left(R_2+x_3+y_3\right)}\bigg]~,\\
G_{23}&=\frac{\left(x_2-y_2\right)}{16 \pi \mu (1-\nu )}\bigg[\frac{x_3-y_3}{{R_1}^3}+\frac{(3-4 \nu ) \left(x_3-y_3\right)}{{R_2}^3} \\
&\qquad\qquad\qquad\qquad-\frac{6\, x_3 y_3 \left(x_3+y_3\right)}{{R_2}^5}+\frac{4 (1-2\nu) (1-\nu )}{R_2\left(R_2+x_3+y_3\right)}\bigg] \\
G_{33}&=\frac{1}{16 \pi\mu(1-\nu )}\bigg[\frac{3-4 \nu }{R_1}+\frac{5-12\nu +8\nu^2}{R_2}+\frac{\left(x_3-y_3\right)^2}{{R_1}^3}\\
&\qquad\qquad\qquad\quad+\frac{6\, x_3 y_3 \left(x_3+y_3\right)^2}{{R_2}^5}+\frac{(3-4\nu)\left(x_3+y_3\right)^2-2\,x_3 y_3}{{R_2}^3}\bigg]~. \\
\end{aligned}
\end{equation}
All involve the radii
\begin{equation}
\begin{aligned}
R_1&=(\left(x_1-y_1\right)^2+\left(x_2-y_2\right)^2+\left(y_3-x_3\right)^2)^{1/2}\\
R_2&=(\left(x_1-y_1\right)^2+\left(x_2-y_2\right)^2+\left(x_3+y_3\right)^2)^{1/2}~.
\end{aligned}
\end{equation}

\subsection*{Semi-analytic solution with the double-exponential and the Gauss-Legendre quadratures}

The integral (\ref{eqn:surface-integral-3d}) involves surface integrals of the form (no summation implied over the indices $i$ and $j$)
\begin{equation}
\begin{aligned}
K_{ij}=\int_{\partial\Omega}&G_{ij}(x_1,x_2,x_3;y_1,y_2,y_3)\left(m_{i1}n_1+m_{i2}n_2+m_{i3}n_3\right)\,\diff y_1\diff y_2\diff y_3~,
\end{aligned}
\end{equation}
that can be broken down into the four faces $ABC$, $BCD$, $CDA$, and $DAB$ of the tetrahedron, as follows
\begin{equation}\label{eqn:list-of-surface-integrals}
\begin{aligned}
K_{ij}
&=\left(m_{i1}n_1^{(D)}+m_{i2}n_2^{(D)}+m_{i3}n_3^{(D)}\right)\int_{ABC}G_{ij}(x_1,x_2,x_3;y_1,y_2,y_3)\,\diff y_1\diff y_2\diff y_3 \\
&+\left(m_{i1}n_1^{(A)}+m_{i2}n_2^{(A)}+m_{i3}n_3^{(A)}\right)\int_{BCD}G_{ij}(x_1,x_2,x_3;y_1,y_2,y_3)\,\diff y_1\diff y_2\diff y_3 \\
&+\left(m_{i1}n_1^{(B)}+m_{i2}n_2^{(B)}+m_{i3}n_3^{(B)}\right)\int_{CDA}G_{ij}(x_1,x_2,x_3;y_1,y_2,y_3)\,\diff y_1\diff y_2\diff y_3 \\
&+\left(m_{i1}n_1^{(C)}+m_{i2}n_2^{(C)}+m_{i3}n_3^{(C)}\right)\int_{DAB}G_{ij}(x_1,x_2,x_3;y_1,y_2,y_3)\,\diff y_1\diff y_2\diff y_3~.
\end{aligned}
\end{equation}
Following the approach described in the previous sections, I obtain solutions to the surface integrals (\ref{eqn:list-of-surface-integrals}) using the double-exponential and the Gauss-Legendre quadratures. To do so, I consider the individual surface integral
\begin{equation}\label{eqn:surface-integral-3d-item}
J_{ij}(x_1,x_2,x_3)=\int_{ABC}G_{ij}(x_1,x_2,x_3;y_1,y_2,y_3)\,\diff y_1\diff y_2\diff y_3~,
\end{equation}
which I write as a parameterized surface integral and in canonical form, to get
\begin{equation}\label{eqn:surface-integral}
J_{ij}(x_1,x_2,x_3)=\frac{\mathcal{A}}{4}\int_{-1}^1\int_{-1}^1(1-v)\,G_{ij}(x_1,x_2,x_3;y_1(u,v),y_2(u,v),y_3(u,v))\,\diff u \diff v~,
\end{equation}
where $\mathcal{A}$ is the area of the triangle ABC and $u$ and $v$ are dummy variables of integration. The parameterization
\begin{equation}
\begin{aligned}
\textbf{y}(u,v)&=\frac{1}{4}\,\textbf{x}^A\,(1-u)\,(1-v) \\
                     &+\frac{1}{4}\,\textbf{x}^B\,(1+u)\,(1-v) \\
                     &+\frac{1}{2}\,\textbf{x}^C\,(1+v)~.
\end{aligned}
\end{equation}
maps the triangle ABC in three-dimensional space to a right isosceles triangle in the $uv$ space~\citep[e.g.,][]{pozrikidis02,beer+08}, where $\textbf{x}^A$, $\textbf{x}^B$, and $\textbf{x}^C$ are the spatial coordinates of vertices A, B, and C in integral (\ref{eqn:surface-integral-3d-item}), respectively.

\begin{figure}[p]
\vspace{-2cm}
\includegraphics[width=\columnwidth]{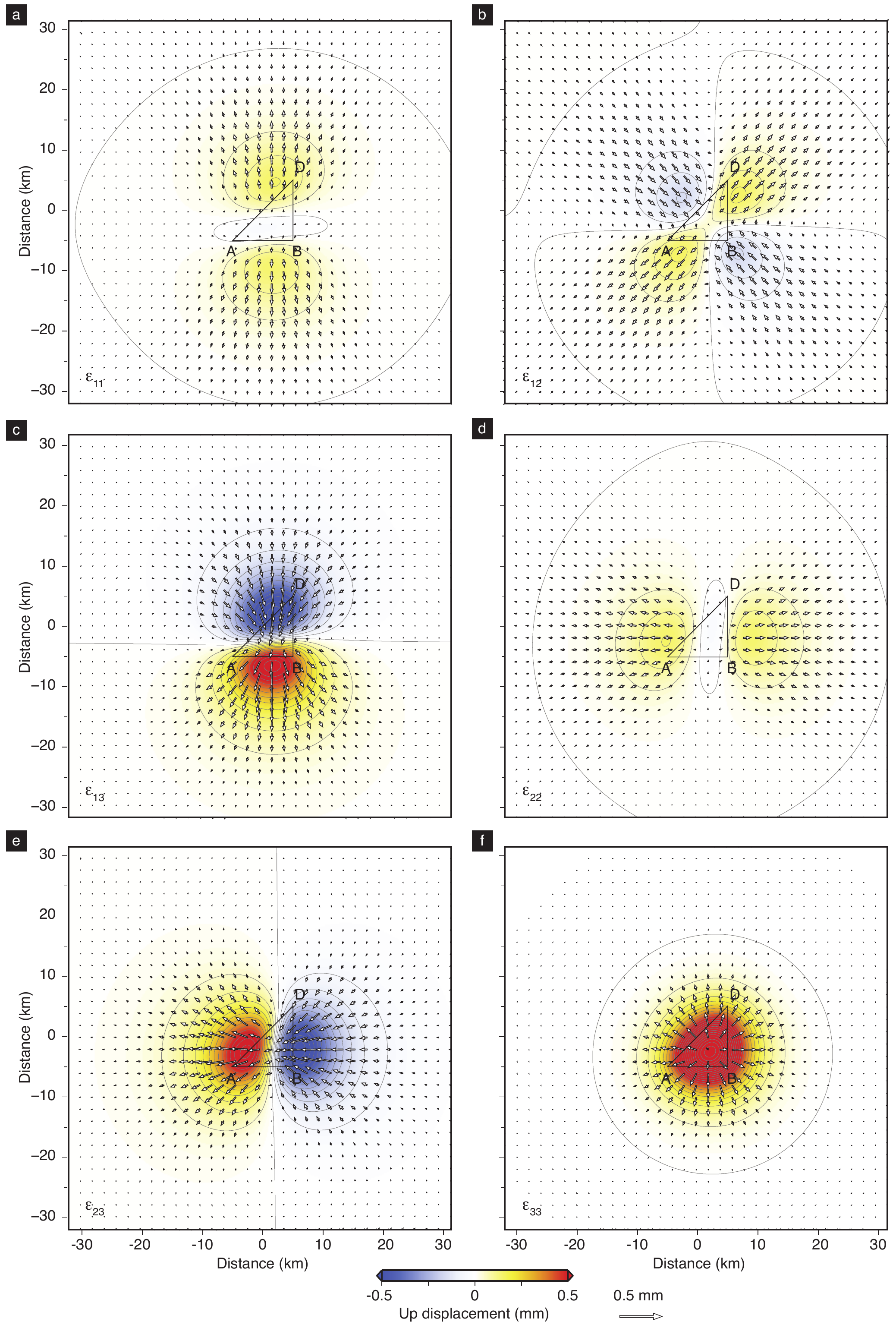}
\caption{\small Displacement field at the surface of the half-space due to anelastic strain confined in a tetrahedral volume ABCD. The arrows indicate horizontal displacements and the background indicates the vertical (positive up) displacement. The panels corresponds to different components of transformation strain with an amplitude of one microstrain: a) uniaxial strain $\epsilon_{11}$, b) pure shear $\epsilon_{12}$, c) vertical pure shear $\epsilon_{13}$, d) horizontal uniaxial extension $\epsilon_{22}$, e) vertical pure shear $\epsilon_{23}$, and f) vertical extension $\epsilon_{33}$. The contours (dashed lines) are every 0.05\,mm.}\label{fig:three-dimension-tetrahedron-eij-uk-map-view}
\end{figure}

For each displacement component correspond twelve integrals such as (\ref{eqn:surface-integral}) due to the presence of three force components on four faces. Therefore, the displacement field requires the evaluation of at most 36 surface integrals. Examples of surface displacements in map view caused by anelastic strain confined in a tetrahedron are shown in Figure~\ref{fig:three-dimension-tetrahedron-eij-uk-map-view}. The vertices are located at $A=(-5,-5,5)$, $B=(-5,5,5)$, $C=(-5,5,15)$, and $D=(5,5,5)$ expressed in km. Each panel shows the displacement caused by a single transformation strain component with an amplitude of one microstrain.

Six tetrahedra can be arranged to form a cuboid. In this case the semi-analytic solution agrees with the analytical solution of~\cite{barbot+17} to the limit of double-precision floating point accuracy. As the numerical solution involves a surface integral, the computational burden is much larger than for the two-dimensional case with only line integrals. The double-exponential quadrature with 601 integration points takes about 2,500 times longer than the analytic solution for a cuboid source. In contrast, the Gauss-Legendre quadrature with 7 and 15 points in both directions takes only 3 times and 16 times longer than the analytic solution, respectively. As the solution based on the Gauss-Legendre quadrature offers the same accuracy but is free of numerical artifacts away from the surface of the tetrahedron, the small difference in computational cost makes this approach more appealing than using the analytic solution.

\subsection*{Stress and strain}

\begin{figure}[p]
\vspace{-1cm}
\includegraphics[width=\columnwidth]{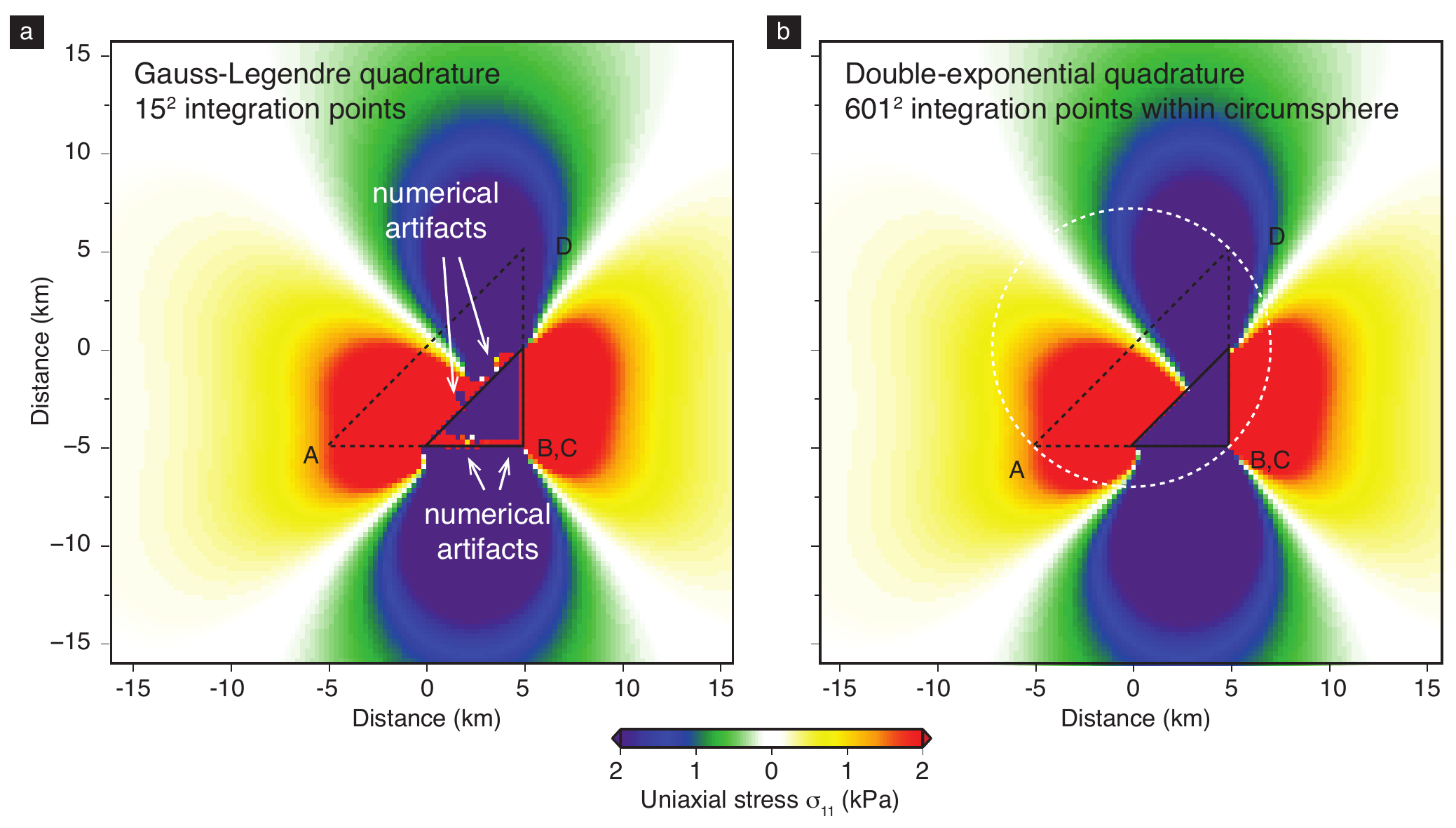}
\caption{\small Uniaxial stress component $\sigma_{11}$ due to non-trivial anelastic strain component $\epsilon_{11}$ evaluated with a) the Gauss-Legendre quadrature throughout the domain with $15^2$ integration points (15 points in each direction of integration of the 4 triangular surfaces) and b) the double-exponential quadrature within the circumsphere (horizontal footprint in dashed white circle) using $601^2$ integration points and the Gauss-Legendre quadrature outside the circumsphere. In both cases, the tetrahedron connects the vertices $A=(-5,-5,5)$, $B=(-5,5,5)$, $C=(-5,5,15)$, and $D=(5,5,5)$ expressed in km (horizontal footprint in dashed black profile) and the figure shows horizontal cross-sections cutting through the tetrahedron at 10\,km depth. The intersection of the cross-section and the tetrahedron is a triangle surface (contour in black profile.) Numerical artifacts near the surface of the tetrahedron are evident with the solution based on the Gauss-Legendre quadrature. They are mostly eliminated with the double-exponential quadrature.}\label{fig:comparison-gauss-legendre-double-exponential-e11-s11-map-view}
\end{figure}

\begin{figure}[p]
\vspace{-3cm}
\includegraphics[width=\columnwidth]{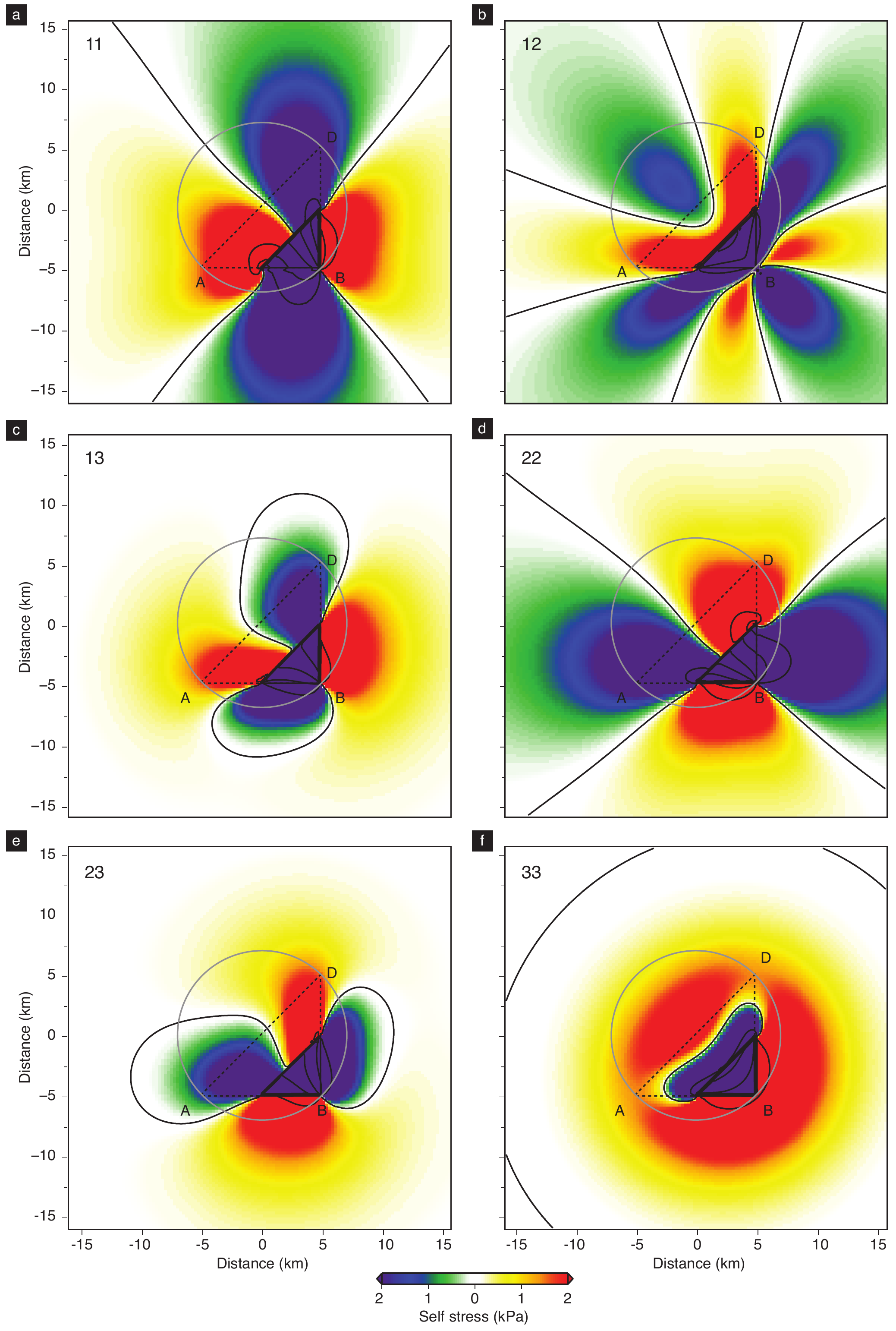}
\caption{\small The stress component $\sigma_{ij}$ in horizontal cross-section due to the nontrivial transformation strain component $\epsilon_{ij}$ confined in a tetrahedron (self-stress). The surface footprint of the tetrahedron is shown in the dashed profile. The intersection of the cross-section with the tetrahedron is shown in solid black profile. The double-exponential quadrature is used for points within the circumsphere (horizontal footprint shown in grey circle) and the Gauss-Legendre quadrature is used outside. The panels correspond to different components of transformation strain with an amplitude of one microstrain: a) uniaxial stress $\sigma_{11}$ due to nontrivial transformation strain component $\epsilon_{11}$, b) stress component $\sigma_{12}$ due to pure shear $\epsilon_{12}$ in the tetrahedron, c) stress component $\sigma_{13}$ due to vertical pure shear $\epsilon_{13}$, d) $\sigma_{22}$ due to horizontal uniaxial extension $\epsilon_{22}$, e) stress component $\sigma_{23}$ due to vertical pure shear $\epsilon_{23}$, and f) vertical stress $\sigma_{33}$ due to vertical extension $\epsilon_{33}$ within the tetrahedron. The contours (dashed lines) are every 10\,kPa.}\label{fig:three-dimension-triangle-eij-sij-map-view}
\end{figure}

\begin{figure}[p]
\vspace{-2cm}
\includegraphics[width=\columnwidth]{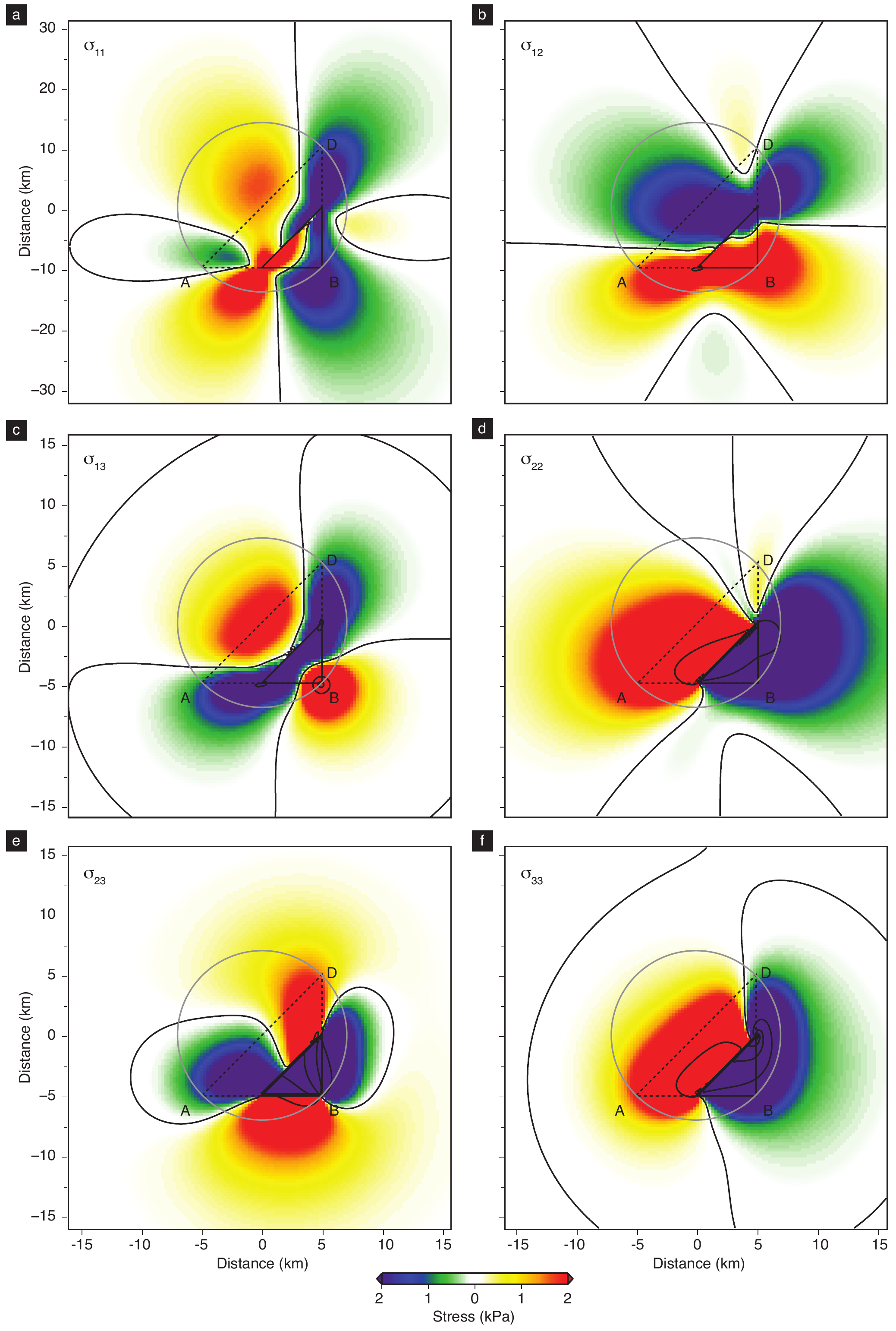}
\caption{\small The stress field in horizontal cross-section due to the nontrivial transformation strain component $\epsilon_{23}$ (one microstrain) confined in a tetrahedron. The surface footprint of the tetrahedron is shown in the dashed profile. The intersection of the cross-section with the tetrahedron is shown in solid black profile. The double-exponential and the Gauss-Legendre quadratures are used for points inside, respectively outside, the circumsphere (horizontal footprint shown in grey circle). a) uniaxial stress $\sigma_{11}$, b) shear stress component $\sigma_{12}$, c) shear stress component $\sigma_{13}$, d) horizontal uniaxial stress component $\sigma_{22}$, e) vertical shear stress component $\sigma_{23}$, and F) vertical stress $\sigma_{33}$. The contours (dashed lines) are every 10\,kPa.}\label{fig:three-dimension-triangle-e23-sij-map-view}
\end{figure}

The stress field is essential to simulate forward models of deformation with the integral method~\citep{barbot18b} and to regularize inverse problems involving distributed strain~\citep{qqiu+18}. The strain can be obtained by differencing the displacement field obtained with~(\ref{eqn:surface-integral-3d}) but a more accurate approach is to directly integrate the Green's functions for the displacement gradient, as in 
\begin{equation}\label{eqn:surface-integral-3d-gradient}
\begin{aligned}
u_{i,j}(x_1,x_2,x_3)=
  \int_{\partial\Omega}&G_{ki,j}(x_1,x_2,x_3;y_1,y_2,y_3)\,m_{kl}\,n_l\,\diff y_1\diff y_2\diff y_3~. \\
\end{aligned}
\end{equation}
The derivatives of the Green's function are given in closed form below for completeness,
\begin{equation}\label{eqn:greens3dstrain1}
\begin{aligned}
G_{11,1}&=\frac{ \left( {x_1}-{y_1} \right)}{16\pi\mu(1-\nu)}\bigg[
-\frac { \left( 3-4\nu \right)}{ {R_1}^3 } -\frac {1}{{R_2}^3} +\frac {  2{R_1}^2-3\left( x_1-y_1 \right) ^2  }{{R_1}^5} \\
&\qquad\qquad\qquad\qquad +\left( 3-4\,\nu \right)\frac {   2{R_2}^2 - 3\left( x_1-y_1 \right) ^2  }{{R_2}^5}  -6\, y_3\,x_3 \frac{3{R_2}^2-5(x_1-y_1)^2}{{R_2}^7} \\
&\qquad\qquad\qquad\qquad -12\frac {  \left( 1-2\nu \right) \left( 1-\nu \right) }{ R_2 \left(R_2+x_3+y_3 \right) ^2}   \\
&\qquad\qquad\qquad\qquad        +\frac{4\,(1-2\nu)(1-\nu)\,(x_1-y_1)^2}{{R_2}^3\left(R_2+x_3+y_3\right)^2}+\frac{8\,(1-2\nu)(1-\nu)\,(x_1-y_1)^2}{{R_2}^2\left(R_2+x_3+y_3\right)^3} \bigg]~,\\
G_{11,2}&=\frac{ \left( {x_2}-{y_2} \right)}{16\pi\mu(1-\nu)}\bigg[
-\frac { \left( 3-4\nu \right)}{ {R_1}^3 } -\frac {1}{{R_2}^3} -\frac {3 \left( x_1-y_1 \right) ^2 }{{R_1}^5}  -\frac {3 \left( 3-4\,\nu \right)  \left( x_1-y_1 \right) ^2 }{{R_2}^5} \\
&\qquad\qquad\qquad\qquad -6\,y_3\,x_3\,\frac{{R_2}^2-5(x_1-y_1)^2}{{R_2}^7}  -\frac { 4\left( 1-2\nu \right) \left( 1-\nu \right) }{ R_2 \left(R_2+x_3+y_3 \right) ^2} \\
&\qquad\qquad\qquad\qquad +4\left( 1-2\nu \right)\left( 1-\nu \right)\left( x_1-y_1 \right)^2\frac {   3R_2+x_3+y_3}{{R_2}^3 \left( R_2+x_3+y_3 \right) ^3} \bigg]~,\\
G_{11,3}&=\frac{ 1 }{16\pi\mu(1-\nu)}\bigg[
-\frac { \left( 3-4\nu \right)\left( {x_3}-{y_3} \right)}{ {R_1}^3 } -\frac {(x_3+y_3)}{{R_2}^3} -3\,\frac { \left( x_1-y_1 \right) ^2  \left( {x_3}-{y_3} \right) }{{R_1}^5} \\
&\qquad\qquad\qquad\qquad -3\,\frac { \left( 3-4\,\nu \right)  \left( x_1-y_1 \right) ^2 (x_3+y_3) }{{R_2}^5} \\
&\qquad\qquad\qquad\qquad +2\,y_3\,\frac{{R_2}^2-3\,x_3(x_3+y_3)}{{R_2}^5} \\
&\qquad\qquad\qquad\qquad -6\,y_3\left(x_1-y_1\right)^2\frac{{R_2}^2-5\,x_3(x_3+y_3)}{{R_2}^7} \\
&\qquad\qquad\qquad\qquad -4\,\frac { \left( 1-2\nu \right) \left( 1-\nu \right) }{ R_2 \left(R_2+x_3+y_3 \right)} \\
&\qquad\qquad\qquad\qquad +4\left( 1-2\nu \right)\left( 1-\nu \right)\left( x_1-y_1 \right)^2\frac {  2R_2+x_3+y_3 }{{R_2}^3 \left( R_2+x_3+y_3 \right) ^2}  \bigg]~,\\
\end{aligned}
\end{equation}
and
\begin{equation}\label{eqn:greens3dstrain2}
\begin{aligned}
G_{21,1}&=\frac{ (x_2-y_2)}{16\pi\mu(1-\nu)}\bigg[
\frac{{R_1}^2-3(x_1-y_1)^2}{{R_1}^5}    + (3-4\nu)\,\frac{{R_2}^2-3(x_1-y_1)^2}{{R_2}^5} \\
&\qquad\qquad\qquad\qquad  - 6\,y_3\,x_3\frac{{R_2}^2-5(x_1-y_1)^2}{{R_2}^7}   - \frac{4(1-2\nu)(1-\nu)}{{R_2}(R_2+x_3+y_3)^2}  \\
&\qquad\qquad\qquad\qquad  + 4\,(1-2\nu)(1-\nu) \left(x_1-y_1\right)^2\frac{3R_2+x_3+y_3}{{R_2}^3(R_2+x_3+y_3)^3}    \bigg]~,\\
G_{21,2}&=\frac{ (x_1-y_1)}{16\pi\mu(1-\nu)}\bigg[
 \frac{{R_1}^2-3(x_2-y_2)^2}{{R_1}^5} + (3-4\nu)\,\frac{{R_2}^2-3(x_2-y_2)^2}{{R_2}^5} \\
&\qquad\qquad\qquad\qquad  - 6\,y_3\,x_3\,\frac{{R_2}^2-5(x_2-y_2)^2}{{R_2}^7}  - \frac{4(1-2\nu)(1-\nu)}{{R_2}(R_2+x_3+y_3)^2}  \\
&\qquad\qquad\qquad\qquad  +4(1-2\nu)(1-\nu) (x_2-y_2)^2\frac{ 3R_2+x_3+y_3 }{{R_2}^3(R_2+x_3+y_3)^3}   \bigg]~,\\
G_{21,3}&=\frac{ \left( {x_1}-{y_1} \right)\left( {x_2}-{y_2} \right)}{16\pi\mu(1-\nu)}\bigg[
-3\,\frac { \left( x_3-y_3\right) }{{R_1}^5}  -3\left( 3-4\nu \right)\frac {  \left( x_3+y_3 \right) }{{R_2}^5} \\
&\qquad\qquad\qquad\qquad\qquad\quad -6\,y_3\frac{{R_2}^2-5\,x_3\,(x_3+y_3)}{{R_2}^7} \\
&\qquad\qquad\qquad\qquad\qquad\quad +4\left( 1-2\nu \right)\left( 1-\nu \right) \frac {   2R_2 + x_3+y_3 }{{R_2}^3 \left( R_2+x_3+y_3 \right) ^2} \bigg]~.\\
\end{aligned}
\end{equation}
Continuing,
\begin{equation}\label{eqn:greens3dstrain3}
\begin{aligned}
G_{31,1}&=\frac{1}{16\pi\mu(1-\nu)}\bigg[
 (x_3-y_3)\frac{{R_1}^2-3(x_1-y_1)^2}{{R_1}^5} \\
&\qquad\qquad\qquad\qquad  + (3-4\nu)(x_3-y_3)\frac{{R_2}^2-3(x_1-y_1)^2}{{R_2}^5} \\
&\qquad\qquad\qquad\qquad +6\,x_3\, y_3\,(x_3+y_3)\frac{{R_2}^2-5(x_1-y_1)^2}{{R_2}^7} \\
&\qquad\qquad\qquad\qquad -4\,(1-2\nu)(1-\nu)\frac{1}{{R_2}({R_2}+x_3+y_3)}   \\
&\qquad\qquad\qquad\qquad +4\,(1-2\nu)(1-\nu)(x_1-y_1)^2\frac{2{R_2}+x_3+y_3}{{R_2}^3({R_2}+x_3+y_3)^2} \bigg]~,\\
G_{31,2}&=\frac{ (x_1-y_1)(x_2-y_2)}{16\pi\mu(1-\nu)}\bigg[
    -3\frac{(x_3-y_3)}{{R_1}^5}      -3(3-4\nu)\frac{(x_3-y_3)}{{R_2}^5}  \\
&\qquad\qquad\qquad     -30 \, y_3\, x_3 \frac{ (x_3+y_3)}{{R_2}^7}      +4(1-2\nu)(1-\nu)\frac{2R_2+x_3+y_3}{{R_2}^3(R_2+x_3+y_3)^2}   \bigg]~,\\
G_{31,3}&=\frac{ \left( {x_1}-{y_1} \right)}{16\pi\mu(1-\nu)}\bigg[
\frac{{R_1}^2-3(x_3-y_3)^2}{{R_1}^5}      +(3-4\nu)\frac{{R_2}^2-3(x_3+y_3)(x_3-y_3)}{{R_2}^5} \\
&\qquad\qquad\qquad     +6\,y_3\frac{{R_2}^2(2x_3+y_3)-5x_3(x_3+y_3)^2}{{R_2}^7}      +4\frac{(1-2\nu)(1-\nu)}{{R_2}^3} 
 \bigg]~.\\
\end{aligned}
\end{equation}
The derivatives of the $G_{12}$ component are the same as for $G_{21}$ component
\begin{equation}\label{eqn:greens3dstrain4}
\begin{aligned}
G_{12,1}&=\frac{ (x_2-y_2)}{16\pi\mu(1-\nu)}\bigg[
\frac{{R_1}^2-3(x_1-y_1)^2}{{R_1}^5}    +(3-4\nu)\frac{{R_2}^2-3(x_1-y_1)^2}{{R_2}^5}  \\
&\qquad   -6\,y_3\,x_3\,\frac{{R_2}^2-5(x_1-y_1)^2}{{R_2}^7}     -\frac{4(1-\nu)(1-2\nu)}{{R_2}(R_2+x_3+y_3)^2} \\
&\qquad    +4(1-\nu)(1-2\nu)(x_1-y_1)^2\frac{3R_2+x_3+y_3}{{R_2}^3(R_2+x_3+y_3)^3}  \bigg]~,\\
G_{12,2}&=\frac{ (x_1-y_1)}{16\pi\mu(1-\nu)}\bigg[
\frac{{R_1}^2-3(x_2-y_2)^2}{{R_1}^5}    +(3-4\nu)\frac{{R_2}^2-3(x_2-y_2)^2}{{R_2}^5} \\
&\qquad    -6\,y_3\,x_3\frac{{R_2}^2-5(x_2-y_2)^2}{{R_2}^7}     -\frac{4(1-\nu)(1-2\nu)}{{R_2}(R_2+x_3+y_3)^2} \\
&\qquad    +4(1-\nu)(1-2\nu)\,(x_2-y_2)^2\frac{3R_2+x_3+y_3}{{R_2}^3(R_2+x_3+y_3)^3} \bigg]~,\\
G_{12,3}&=\frac{ \left( {x_1}-{y_1} \right)\left( {x_2}-{y_2} \right)}{16\pi\mu(1-\nu)}\bigg[
-3\,\frac { \left( x_3-y_3\right) }{{R_1}^5}  -3\left( 3-4\nu \right)\frac {  \left( x_3+y_3 \right) }{{R_2}^5} \\
&\qquad\qquad\qquad\qquad\qquad\quad -6\,y_3\frac{{R_2}^2-5\,x_3\,(x_3+y_3)}{{R_2}^7} \\
&\qquad\qquad\qquad\qquad\qquad\quad +4\left( 1-2\nu \right)  \left( 1-\nu \right)\frac {   2R_2+x_3+y_3 }{{R_2}^3 \left( R_2+x_3+y_3 \right) ^2} \bigg]~.\\
\end{aligned}
\end{equation}
The derivatives of the $G_{22}$ have a symmetry with those of the $G_{11}$ component by permutation of the $1$ and $2$ indices
\begin{equation}\label{eqn:greens3dstrain5}
\begin{aligned}
G_{22,1}&=\frac{ \left( {x_1}-{y_1} \right)}{16\pi\mu(1-\nu)}\bigg[
-\frac { \left( 3-4\nu \right)}{ {R_1}^3 } -\frac {1}{{R_2}^3} -\frac {3 \left( x_2-y_2 \right) ^2 }{{R_1}^5}  -\frac {3 \left( 3-4\,\nu \right)  \left( x_2-y_2 \right) ^2 }{{R_2}^5} \\
&\qquad\qquad -6\,y_3\,x_3 \,\frac{{R_2}^2-5(x_2-y_2)^2}{{R_2}^7}  -\frac {  4\left( 1-2\nu \right) \left( 1-\nu \right) }{ R_2 \left(R_2+x_3+y_3 \right) ^2} \\
&\qquad\qquad + 4\left( 1-2\nu \right)\left( 1-\nu \right) \left( x_2-y_2 \right)^2\frac {   3R_2+x_3+y_3 }{{R_2}^3 \left( R_2+x_3+y_3 \right) ^3} \bigg]~,\\
G_{22,2}&=\frac{ \left( {x_2}-{y_2} \right)}{16\pi\mu(1-\nu)}\bigg[
-\frac { \left( 3-4\nu \right)}{ {R_1}^3 } -\frac {1}{{R_2}^3} + \frac {2{R_1}^2 - 3\left( x_2-y_2 \right) ^2 }{{R_1}^5} \\
&\qquad\qquad\qquad\qquad + \left( 3-4\,\nu \right)\frac {  2{R_2}^2-3\left( x_2-y_2 \right) ^2  }{{R_2}^5}  - 6\,y_3 \, x_3\frac{3{R_2}^2-5(x_2-y_2)^2}{{R_2}^7} \\
&\qquad\qquad\qquad\qquad -12 \frac { \left( 1-2\nu \right) \left( 1-\nu \right) }{ {R_2} \left(R_2+x_3+y_3 \right) ^2} \\
&\qquad\qquad\qquad\qquad +4 \left( 1-2\nu \right) \left( 1-\nu \right)(x_2-y_2)^2\frac { 3R_2+x_3+y_3}{ {R_2}^3 \left(R_2+x_3+y_3 \right) ^3}  \bigg]~,\\
G_{22,3}&=\frac{ 1 }{16\pi\mu(1-\nu)}\bigg[
-\left( 3-4\nu \right)\frac { \left( {x_3}-{y_3} \right)}{ {R_1}^3 } -\frac {(x_3+y_3)}{{R_2}^3} -3\,\frac { \left( x_2-y_2 \right) ^2  \left( {x_3}-{y_3} \right) }{{R_1}^5} \\
&\qquad\qquad\qquad\qquad -3\left( 3-4\,\nu \right) \frac {  \left( x_2-y_2 \right) ^2 (x_3+y_3) }{{R_2}^5} \\
&\qquad\qquad\qquad\qquad +2\,y_3\,\frac{{R_2}^2-3\,x_3(x_3+y_3)}{{R_2}^5} \\
&\qquad\qquad\qquad\qquad -6\,y_3\left(x_2-y_2\right)^2\frac{{R_2}^2-5\,x_3(x_3+y_3)}{{R_2}^7} \\
&\qquad\qquad\qquad\qquad -4\frac { \left( 1-2\nu \right) \left( 1-\nu \right) }{ R_2 \left(R_2+x_3+y_3 \right)} \\
&\qquad\qquad\qquad\qquad +4\left( 1-2\nu \right)\left( 1-\nu \right)\left( x_2-y_2 \right)^2\frac {   2R_2+x_3+y_3 }{{R_2}^3 \left( R_2+x_3+y_3 \right) ^2}  \bigg]~,\\
\end{aligned}
\end{equation}
The derivatives of the $G_{32}$ term can be obtained from the $G_{31}$ term by permutation of the 1 and 2 indices,
\begin{equation}\label{eqn:greens3dstrain6}
\begin{aligned}
G_{32,1}&=\frac{ (x_1-y_1)(x_2-y_2)}{16\pi\mu(1-\nu)}\bigg[
    -3\frac{(x_3-y_3)}{{R_1}^5}      -3(3-4\nu)\frac{(x_3-y_3)}{{R_2}^5}  \\
&\qquad\qquad\qquad     -30 \,y_3\, x_3 \frac{ (x_3+y_3)}{{R_2}^7}      +4(1-2\nu)(1-\nu)\frac{2R_2+x_3+y_3}{{R_2}^3(R_2+x_3+y_3)^2}   \bigg]~,\\
G_{32,2}&=\frac{1}{16\pi\mu(1-\nu)}\bigg[
 (x_3-y_3)\frac{{R_1}^2-3(x_2-y_2)^2}{{R_1}^5} \\
&\qquad\qquad\qquad  + (3-4\nu)(x_3-y_3)\frac{{R_2}^2-3(x_2-y_2)^2}{{R_2}^5} \\
&\qquad\qquad\qquad +6\, y_3\,x_3 \, (x_3+y_3)\frac{{R_2}^2-5(x_2-y_2)^2}{{R_2}^7} \\
&\qquad\qquad\qquad -4\left(1-2\nu\right)\left(1-\nu\right)\frac{ 1 }{{R_2}({R_2}+x_3+y_3)}  \\
&\qquad\qquad\qquad +4\left(1-2\nu\right)\left(1-\nu\right)(x_2-y_2)^2\frac{2{R_2}+x_3+y_3}{{R_2}^3({R_2}+x_3+y_3)^2} \bigg]~,\\
G_{32,3}&=\frac{ \left( {x_2}-{y_2} \right)}{16\pi\mu(1-\nu)}\bigg[
\frac{{R_1}^2-3(x_3-y_3)^2}{{R_1}^5}      +(3-4\nu)\frac{{R_2}^2-3(x_3-y_3)(x_3+y_3)}{{R_2}^5} \\
&\qquad\qquad\qquad     +6\,y_3\frac{(2x_3+y_3)}{{R_2}^5} -30\,y_3\,x_3\frac{(x_3+y_3)^2}{{R_2}^7}      +4\frac{(1-2\nu)(1-\nu)}{{R_2}^3} 
 \bigg]~.\\
\end{aligned}
\end{equation}
The derivatives of the $G_{13}$ component are
\begin{equation}\label{eqn:greens3dstrain7}
\begin{aligned}
G_{13,1}&=\frac{ 1}{16\pi\mu(1-\nu)}\bigg[
\left(x_3-y_3\right)\frac{{R_1}^2-3(x_1-y_1)^2}{{R_1}^5} \\
& +(3-4\nu)\left(x_3-y_3\right)\frac{{R_2}^2-3(x_1-y_1)^2}{{R_2}^5} \\ 
& -6\,y_3\,x_3\left(x_3+y_3\right)\frac{{R_2}^2-5(x_1-y_1)^2}{{R_2}^7}  +\frac{4(1-2\nu)(1-\nu)}{{R_2}(R_2+x_3+y_3)} \\
& -4(1-2\nu)(1-\nu)(x_1-y_1)^2\frac{2{R_2}+x_3+y_3}{{R_2}^3(R_2+x_3+y_3)^2}  \bigg]~,\\
G_{13,2}&=\frac{(x_1-y_1)(x_2-y_2)}{16\pi\mu(1-\nu)}\bigg[
-3(x_3-y_3)\frac{1}{{R_1}^5} \\
& -3\, (3-4\nu)\,(x_3-y_3)\frac{1}{{R_2}^5} \\
& +30\, y_3\,x_3\,(x_3+y_3)\frac{1}{{R_2}^7} \\
& -4(1-2\nu)(1-\nu)\frac{2{R_2}+x_3+y_3}{{R_2}^3({R_2}+x_3+y_3)^2}
 \bigg]~,\\
G_{13,3}&=\frac{ \left( {x_1}-{y_1} \right)}{16\pi\mu(1-\nu)}\bigg[
\frac{{R_1}^2-3(x_3-y_3)^2}{{R_1}^5} \\
& +(3-4\nu) \frac{{R_2}^2-3(x_3-y_3)(x_3+y_3)}{{R_2}^5} \\
& -6\,y_3\frac{2x_3+y_3}{{R_2}^5}  +30\,y_3\,x_3\frac{(x_3+y_3)^2}{{R_2}^7} \\
& -4 (1-2\nu)(1-\nu)\frac{1}{{R_2}^3}  \bigg]~.\\
\end{aligned}
\end{equation}
The derivatives of the $G_{23}$ can be obtained from the $G_{13}$ derivatives by permutation of the $1$ and $2$ indices
\begin{equation}\label{eqn:greens3dstrain8}
\begin{aligned}
G_{23,1}&=\frac{(x_1-y_1)(x_2-y_2)}{16\pi\mu(1-\nu)}\bigg[
-3(x_3-y_3)\frac{1}{{R_1}^5} \\
& -3\, (3-4\nu)\,(x_3-y_3)\frac{1}{{R_2}^5} \\
& +30\, y_3\,x_3\,(x_3+y_3)\frac{1}{{R_2}^7} \\
& -4(1-2\nu)(1-\nu)\frac{2{R_2}+x_3+y_3}{{R_2}^3({R_2}+x_3+y_3)^2}
 \bigg]~,\\
G_{23,2}&=\frac{ 1}{16\pi\mu(1-\nu)}\bigg[
\left(x_3-y_3\right)\frac{{R_1}^2-3(x_2-y_2)^2}{{R_1}^5} \\
& +(3-4\nu)\left(x_3-y_3\right)\frac{{R_2}^2-3(x_2-y_2)^2}{{R_2}^5} \\ 
& -6\,y_3\,x_3\left(x_3+y_3\right)\frac{{R_2}^2-5(x_2-y_2)^2}{{R_2}^7}  +\frac{4(1-2\nu)(1-\nu)}{{R_2}(R_2+x_3+y_3)} \\
& -4(1-2\nu)(1-\nu)(x_2-y_2)^2\frac{2{R_2}+x_3+y_3}{{R_2}^3(R_2+x_3+y_3)^2} \bigg]~,\\
G_{23,3}&=\frac{ \left( {x_2}-{y_2} \right)}{16\pi\mu(1-\nu)}\bigg[
 \frac{{R_1}^2-3(x_3-y_3)^2}{{R_1}^5} \\
& +(3-4\nu) \frac{{R_2}^2-3(x_3-y_3)(x_3+y_3)}{{R_2}^5} \\
& -6\,y_3\frac{2x_3+y_3}{{R_2}^5}  +6\,y_3\,x_3\frac{(x_3+y_3)^2}{{R_2}^7} \\
& -4 (1-2\nu)(1-\nu)\frac{1}{{R_2}^3}  \bigg]~.\\
\end{aligned}
\end{equation}
Finally, the derivatives of the $G_{33}$ component are
\begin{equation}\label{eqn:greens3dstrain8}
\begin{aligned}
G_{33,1}&=\frac{ (x_1-y_1) }{16\pi\mu(1-\nu)}\bigg[
-(3-4\nu)\frac{1}{{R_1}^3} \\
& -(5-12\nu+8\nu^2)\frac{1}{{R_2}^3} \\
& -3\,\frac{(x_3-y_3)^2}{{R_1}^5}  -30\,  y_3\, x_3 \frac{(x_3+y_3)^2}{{R_2}^7} \\
& -3 (3-4\nu) \frac{(x_3+y_3)^2}{{R_2}^5}  + 6\frac{y_3x_3}{{R_2}^5}
 \bigg]~,\\
G_{33,2}&=\frac{ (x_2-y_2) }{16\pi\mu(1-\nu)}\bigg[
-(3-4\nu)\frac{1}{{R_1}^3} \\
& -(5-12\nu+8\nu^2)\frac{1}{{R_2}^3} \\
& -3\,\frac{(x_3-y_3)^2}{{R_1}^5}  -30\,  y_3\, x_3 \frac{(x_3+y_3)^2}{{R_2}^7} \\
& -3 (3-4\nu) \frac{(x_3+y_3)^2}{{R_2}^5}  + 6\frac{y_3x_3}{{R_2}^5}
 \bigg]~,\\
G_{33,3}&=\frac{ 1 }{16\pi\mu(1-\nu)}\bigg[
-(3-4\nu)\frac{\left( {x_3}-{y_3} \right)}{{R_1}^3} \\
& -(5-12\nu+8\nu^2)\frac{\left( {x_3}+{y_3} \right)}{{R_2}^3} \\
& +(x_3-y_3)\frac{2{R_1}^2-3(x_3-y_3)^2}{{R_1}^5}  +6\,y_3\,\frac{(x_3+y_3)^2}{{R_2}^5} \\
& +6\,y_3\,x_3\,(x_3+y_3)\frac{2{R_2}^2-5(x_3+y_3)^2}{{R_2}^7} \\
& +(3-4\nu)\,(x_3+y_3)\frac{2{R_2}^2-3(x_3+y_3)^2}{{R_2}^5} \\
& -2\,y_3\frac{{R_2}^2-3x_3(x_3+y_3)}{{R_2}^5} \bigg]~.\\
\end{aligned}
\end{equation}
In all the terms composing the Green's functions (\ref{eqn:greens3d1}-\ref{eqn:greens3d3}) and their derivatives (\ref{eqn:greens3dstrain1}-\ref{eqn:greens3dstrain8}), the only singular point is at $\textbf{x}=\textbf{y}$, guaranteeing that the numerical integration of the Green's functions or their derivatives will be numerically stable at any point away from the surface compounding the source.

For a tetrahedral region with vertices A, B, C, and D as in Figure~\ref{fig:geometry-3d-tetrahedron}, the location of the transformation strain is given by the product of four Heaviside functions
\begin{equation}\label{eqn:omega-tetrahedron}
\begin{aligned}
\Phi(\textbf{x})=
&H\left[\left(\frac{\textbf{x}^A+\textbf{x}^B+\textbf{x}^C}{3}-\textbf{x}\right)\cdot\textbf{n}^{(D)}\right] \\
\times\,&H\left[\left(\frac{\textbf{x}^B+\textbf{x}^C+\textbf{x}^D}{3}-\textbf{x}\right)\cdot\textbf{n}^{(A)}\right] \\
\times\,&H\left[\left(\frac{\textbf{x}^C+\textbf{x}^D+\textbf{x}^A}{3}-\textbf{x}\right)\cdot\textbf{n}^{(B)}\right] \\
\times\,&H\left[\left(\frac{\textbf{x}^D+\textbf{x}^A+\textbf{x}^B}{3}-\textbf{x}\right)\cdot\textbf{n}^{(C)}\right]~.
\end{aligned}
\end{equation}
The stress field can then be obtained using (\ref{eqn:displacement-strain}), (\ref{eqn:stress-eigenstrain}), (\ref{eqn:surface-integral-3d-gradient}), and (\ref{eqn:omega-tetrahedron}). 

The numerical solutions for the stress field based on the Gauss-Legendre and the double-exponential quadratures are compared in Figure~\ref{fig:comparison-gauss-legendre-double-exponential-e11-s11-map-view} for the case of a tetrahedron with the vertices $A=(-5,-5,5)$, $B=(-5,5,5)$, $C=(-5,5,15)$, and $D=(5,5,5)$ expressed in km and a non-trivial transformation strain $\epsilon_{11}$ of one microstrain. Using 15 integration points in each direction of integration with the Gauss-Legendre quadrature, some numerical artifacts scatter in the near-field, close to the surface of the tetrahedron. A close inspection of the residuals with analytic solutions shows that the numerical error decays away from the surface with a radial dependence. In the far-field, a low-order quadrature is sufficient to obtain double-precision accuracy, as noted by~\cite{segall10}. In the near-field, the error can can be eliminated with the double-exponential quadrature using more integration points. With 601 integration points in both directions of integration, the errors are less than can be represented with double-precision arithmetics. 

Based on these results a simple heuristic can be divised to eliminate numerical errors and minimize the computational cost of these calculations whereby the double-exponential or the Gauss-Legendre quadrature is used depending on the distance from the center of the circumsphere. This approach guarantees double-precision accuracy with a computational cost comparable to using analytic solutions, all the while avoiding all possible numerical artifacts away from the surface of the tetrahedron. Some examples of stress interactions are shown in Figures~\ref{fig:three-dimension-triangle-eij-sij-map-view} and~\ref{fig:three-dimension-triangle-e23-sij-map-view} where the double-exponential quadrature was used for points within one radius from the circumsphere center.

\section*{Semi-analytic solution with a spectral method}

\begin{figure}
\includegraphics[width=\columnwidth]{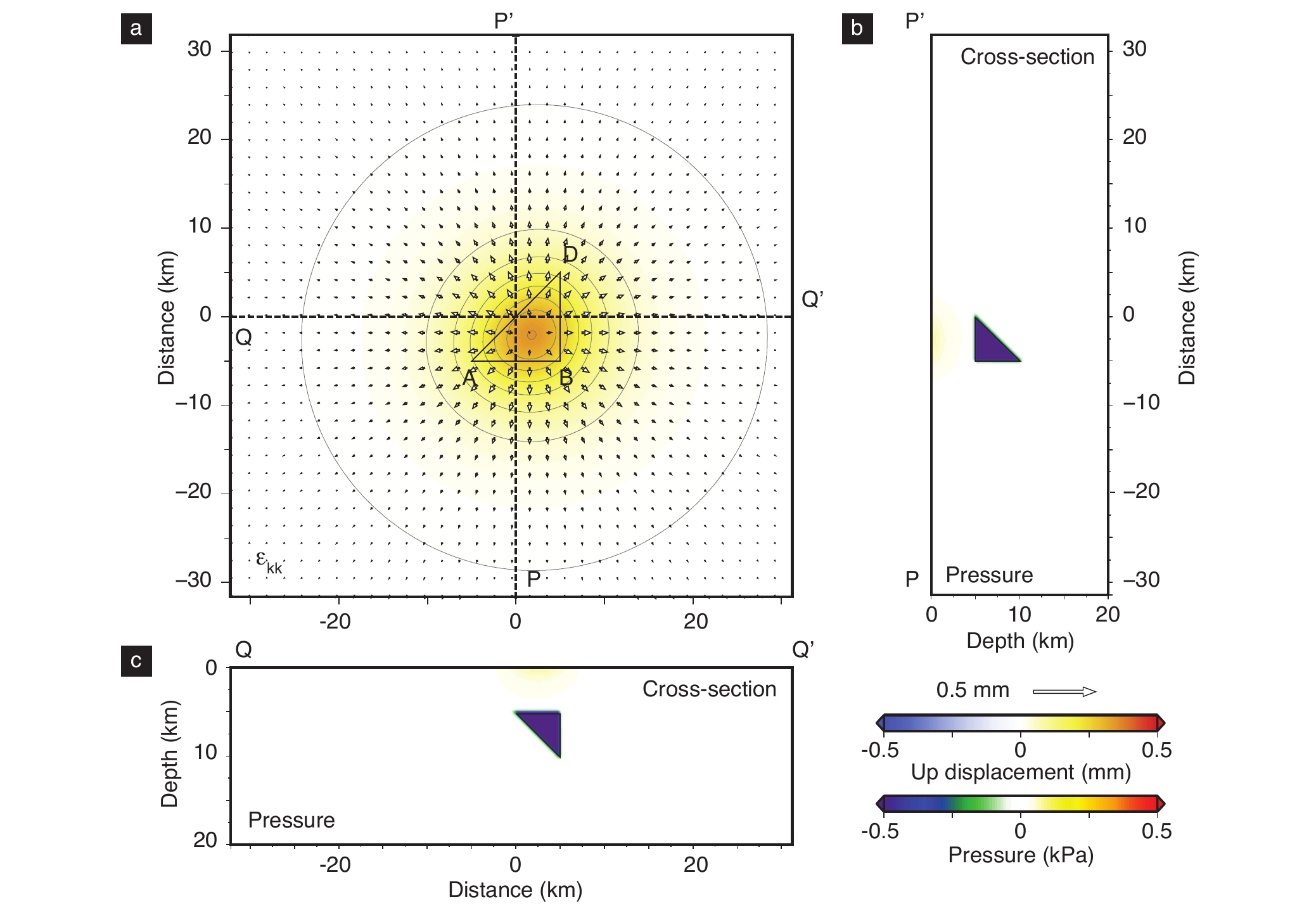}
\caption{\small Displacement at the surface and pressure in the half-space due to isotropic transformation strain confined in a tetrahedral volume ABCD calculated with a spectral method. A) The arrows indicate horizontal displacements and the background indicates the vertical (positive up) displacement. B) The pressure field in the cross-section P-P'. C) The pressure in the cross-section Q-Q'. The transformation strain is $\epsilon_{11}=\epsilon_{22}=\epsilon_{33}=0.33\times10^{-6}$. The vertical displacement contours are every 0.05\,mm.}\label{fig:relax-tetrahedron-ekk-p}
\end{figure}

I now derive numerical solutions compatible with the Fourier-domain semi-analytic solver of~\cite{barbot+fialko10a} implemented in the software \textit{Relax} that was recently optimized for parallel computing on GPU~\citep{masuti+14}. The approach solves Navier's equation (\ref{eqn:momentum}) analytically in the Fourier domain and provides the space-domain solution using a discrete Fourier transform. This approach is numerically more efficient than employing an analytic solution for large enough domains because of the scaling properties of the fast Fourier transform~\cite[e.g.,][]{liu+wang05,sliu+12}. However, the numerical accuracy is limited to about 1\% due to undesirable periodic boundary conditions. At the heart of the method is the explicit sampling of the equivalent body-force density~(\ref{eqn:eqbf-simplified}). This is obtained with
\begin{equation}\label{eqn:grad-omega-tetrahedron}
\begin{aligned}
-\nabla\Phi(\textbf{x})=
\textbf{n}^{(D)}\,\delta&\left[\left(\frac{\textbf{x}^A+\textbf{x}^B+\textbf{x}^C}{3}-\textbf{x}\right)\cdot\textbf{n}^{(D)}\right]\,.\,H\left[\left(\frac{\textbf{x}^B+\textbf{x}^C+\textbf{x}^D}{3}-\textbf{x}\right)\cdot\textbf{n}^{(A)}\right] \\
.\,H&\left[\left(\frac{\textbf{x}^C+\textbf{x}^D+\textbf{x}^A}{3}-\textbf{x}\right)\cdot\textbf{n}^{(B)}\right]\,.\,H\left[\left(\frac{\textbf{x}^D+\textbf{x}^A+\textbf{x}^B}{3}-\textbf{x}\right)\cdot\textbf{n}^{(C)}\right] \\
+~\textbf{n}^{(A)}\,H&\left[\left(\frac{\textbf{x}^A+\textbf{x}^B+\textbf{x}^C}{3}-\textbf{x}\right)\cdot\textbf{n}^{(D)}\right]\,.\,\delta\left[\left(\frac{\textbf{x}^B+\textbf{x}^C+\textbf{x}^D}{3}-\textbf{x}\right)\cdot\textbf{n}^{(A)}\right] \\
.\,H&\left[\left(\frac{\textbf{x}^C+\textbf{x}^D+\textbf{x}^A}{3}-\textbf{x}\right)\cdot\textbf{n}^{(B)}\right]\,.\,H\left[\left(\frac{\textbf{x}^D+\textbf{x}^A+\textbf{x}^B}{3}-\textbf{x}\right)\cdot\textbf{n}^{(C)}\right] \\
+~\textbf{n}^{(B)}\,H&\left[\left(\frac{\textbf{x}^A+\textbf{x}^B+\textbf{x}^C}{3}-\textbf{x}\right)\cdot\textbf{n}^{(D)}\right]\,.\,H\left[\left(\frac{\textbf{x}^B+\textbf{x}^C+\textbf{x}^D}{3}-\textbf{x}\right)\cdot\textbf{n}^{(A)}\right] \\
.\,\delta&\left[\left(\frac{\textbf{x}^C+\textbf{x}^D+\textbf{x}^A}{3}-\textbf{x}\right)\cdot\textbf{n}^{(B)}\right]\,.\,H\left[\left(\frac{\textbf{x}^D+\textbf{x}^A+\textbf{x}^B}{3}-\textbf{x}\right)\cdot\textbf{n}^{(C)}\right] \\
+~\textbf{n}^{(C)}\,H&\left[\left(\frac{\textbf{x}^A+\textbf{x}^B+\textbf{x}^C}{3}-\textbf{x}\right)\cdot\textbf{n}^{(D)}\right]\,.\,H\left[\left(\frac{\textbf{x}^B+\textbf{x}^C+\textbf{x}^D}{3}-\textbf{x}\right)\cdot\textbf{n}^{(A)}\right] \\
.\,H&\left[\left(\frac{\textbf{x}^C+\textbf{x}^D+\textbf{x}^A}{3}-\textbf{x}\right)\cdot\textbf{n}^{(B)}\right]\,.\,\delta\left[\left(\frac{\textbf{x}^D+\textbf{x}^A+\textbf{x}^B}{3}-\textbf{x}\right)\cdot\textbf{n}^{(C)}\right]~.
\end{aligned}
\end{equation}
For the stability of the Fourier transform and to avoid Gibbs oscillations near sharp discontinuities, the Heaviside function is replaced with an error function
\begin{equation}\label{eqn:H}
H(x)\sim\frac{1}{2}\left[1+\textrm{erf}\left(\frac{x}{\sigma\sqrt{2}}\right)\right]
\end{equation}
and the Delta function is replaced with a Gaussian function
\begin{equation}\label{eqn:D}
\delta(x)\sim\frac{1}{\sqrt{2\pi\sigma^2}}\textrm{exp}\left(-\frac{x^2}{2\sigma^2}\right)
\end{equation}
Both approximations are exact in the limit $\sigma\rightarrow0$ and in practice I employ $\sigma=\Delta x$, using the numerical sampling size as a smoothing factor. This choice is appropriate to conserve linear momentum, i.e., (\ref{eqn:D}) is the derivative of (\ref{eqn:H}), and to suppress singular points near the vertices of the tetrahedron. Figure~\ref{fig:relax-tetrahedron-ekk-p} shows the displacement field and the pressure field induced by an isotropic transformation strain, which may find some applications in hydrological studies. The isotropic strain does not impact a change of pressure in the surrounding medium except near the free surface, as remarked earlier~\citep{faivre69,barbot+17}. This observation also serves as a sophisticated benchmark.

\section*{Conclusions}

I have presented solutions for the displacement and stress kernels in a half-space associated with transformation strain confined in a tetrahedral volume. Numerical and analytic solutions provide the same accuracy, can be evaluated with commensurate computational costs, but numerical quadratures can be more stable in some cases. This work may afford more accurate models of distributed deformation that incorporate structural data. While I hope these results will be useful, some key elements are missing, such as the stratification of elastic properties, surface topography~\citep[e.g.,][]{mayo85,mctigue&segall88,mckenney+95,cayol+cornet97,williams+wadge98,williams+wadge00,wang+18}, in particular, Earth's curvature~\citep{pollitz97a,yu+okubo16}, and coupling with gravity~\citep[][]{rundle82,okubo92,wang+06a}. For more realistic models with laterial variations of elastic moduli, fully numerical methods may be employed~\citep[e.g.,][]{landry+barbot16,landry+barbot18}.

\section*{Acknowledgements}
The author gratefully acknowledges the thoughtful comments of three anonymous reviewers. This research was supported by the National Research Foundation of Singapore under the NRF Fellowship scheme (National Research Fellow Awards No. NRF-NRFF2013-04) and by the Earth Observatory of Singapore, the National Research Foundation, and the Singapore Ministry of Education under the Research Centres of Excellence initiative.

\section*{Data and Resources}
The MATLAB computer programs used in the manuscript are available at \textit{https://bitbucket.org/sbarbot} (last accessed in June 2018). The \textit{Relax} modeling software is hosted at \textit{www.geodynamics.org} (last accessed in February 2018) with support from the Computational Infrastructure for Geodynamics.

\renewcommand\refname{References}

\clearpage

\section*{Appendix A. Analytic solution for line integrals of the half-space elasto-static Green's function in plane strain}

\setcounter{equation}{0}
\renewcommand{\theequation}{A\arabic{equation}}

The closed-form solutions for the line integral (\ref{eqn:closed-form-in-plane-strain}) of the half-space Green's functions can be obtained using symbolic algebra with \textit{Maple}$^{\textregistered}$. The displacement field only depends on the coordinates of the end-points of the line integral, so that 
\begin{equation}
U_{ij}=\int_{AB}G_{ij}(x_2,x_3;y_2,y_3)\,\diff y_2\diff y_3=I_{ij}\left(\frac{R}{2}\right)-I_{ij}\left(-\frac{R}{2}\right)~,
\end{equation}
for $i,j=2,3$, where $R$ is the length of segment AB. For increased clarity, I define the mid-point coordinates
\begin{equation}
\textbf{m}=\frac{\textbf{x}^A+\textbf{x}^B}{2}
\end{equation}
and the azimuthal vector
\begin{equation}
\textbf{a}=\frac{\textbf{x}^B-\textbf{x}^A}{||AB||}
\end{equation}
The expressions for $I_{ij}(t)$ are provided below. 
\normalsize
\begin{align*}
I_{22}(t)& = \\
&\frac{-1}{2\pi \left( -1+\nu \right)} \bigg\{  \bigg( \frac{1}{2}\,t \left( -3/4+\nu \right) {{ a_2}}^{4}+1/2\, \left( -3/4+\nu \right) \left( -{ x_2}+{ m_2} \right) {{ a_2}}^{3} \\
&\qquad\qquad  -1/2\, \left( -2\,t \left(-3/4+\nu \right) { a_3}+ \left( \nu-5/4 \right)  \left( { x_3}-{ m_3} \right)  \right) { a_3}\,{{ a_2}}^{2} \\
&\qquad\qquad+1/2\, \left( -1/4+\nu \right) {{ a_3}}^{2} \left( -{
 x_2}+{ m_2} \right) { a_2}-1/2\, \left( { x_3}-{ m_3}-{ a_3}\,t \right) {{ a_3}}^{3} \left( -3/4+\nu \right)  \bigg) \\
&\qquad\qquad . \ln  \bigg[   {t}^{2}+ \left(  \left( 2\,{ m_2}-2\,{ x_2} \right) { a_2}+2\,{ a_3}\, \left( -{ x_3}+{ m_3} \right)  \right) t+(x_2-m_2)^2+(x_3-m_3)^2 \bigg] \\
&\qquad\qquad+ \left(  \left( { x_3}-{ m_3} \right) { a_2}+{ a_3}\, \left( -{ x_2}+{ m_2} \right)  \right) \left(  \left( -1+\nu \right) {{ a_2}}^{2}+ \left( -1/2+\nu \right) {{ a_3}}^{2} \right)  \\
& \qquad\qquad\qquad . \arctan \left[ {\frac {-t{{ a_2}}^{2}+ \left( -{ m_2}+{
 x_2} \right) { a_2}-{ a_3}\, \left( { a_3}\,t-{ x_3}+{ m_3} \right) }{ \left( -{ x_3}+{ m_3} \right) { a_2}-{ a_3}\, \left( -{ x_2}+{ m_2} \right) }}
 \right] \\
&\qquad\qquad- \left(  \left( -3/4+\nu \right) {{ a_2}}^{2}+ \left( -1/2+\nu \right) {{ a_3}}^{2} \right) t   ~ \bigg\}   \\
&+\frac{5-12\,\nu+8{\nu}^{2} }{8\pi\left( -1+\nu \right) }  \bigg\{  \big( 1/2\,t{{ a_2}}^{2}+ \left( 1/2\,{ m_2}-1/2\,{ x_2} \right) { a_2}+1/2\,{ a_3}\, \left( { x_3}+{ m_3}+{
 a_3}\,t \right)  \big) \\
&\qquad\qquad . \ln  \bigg[   {t}^{2}+ \left(  \left( 2\,{ m_2}-2\,{ x_2} \right) { a_2}+2\,{ a_3}\, \left( {
 x_3}+{ m_3} \right)  \right) t+(x_2-m_2)^2+(x_3+m_3)^2 \bigg] \\
 &\qquad\qquad+ \left(  \left( { x_3}+
{ m_3} \right) { a_2}-{ a_3}\, \left( -{ x_2}+{ m_2} \right)  \right) \\
&\qquad\qquad . \arctan \left[ {\frac {t{{ a_2}}^{2}+ \left( -{ x_2}+{ m_2} \right) { a_2}+{ a_3}\, \left( { x_3}+{ m_3}+{ a_3}\,t \right) }{ \left( { x_3}+{ m_3} \right) { a_2}-{ a_3}\, \left( -{ x_2}+{ m_2} \right) }} \right] -  t~ \bigg\}  \\
&+\frac{1}{8\pi\left( -1+\nu \right)}\, \bigg[ -4\, \big(  \left(  \left( { x_3}+{ m_3} \right) \nu-{ x_3}-3/4\,{ m_3} \right) {{ a_2}}^{2} \\
&\qquad\qquad\qquad- \left( -3/4+\nu \right) { a_3}\, \left( -{ x_2}+{
 m_2} \right) { a_2}-1/4\,{{ a_3}}^{2}{ x_3} \big)  \big( { a_3}\, \left( -{ m_2}+{ x_2} \right) + \left( { x_3}+{ m_3} \right) { a_2} \big) { a_3}
\, \\
&\qquad\qquad . \ln  \bigg[   {t}^{2}+ \left(  \left( 2\,{ m_2}-2\,{ x_2} \right) { a_2}+2\,{ a_3}\, \left( { x_3}+{ m_3} \right)
 \right) t+(x_2-m_2)^2+(x_3+m_3)^2 \bigg] \\
 &\qquad\qquad + \bigg\{  \big( 4\, \left( -{ x_2}+{ m_2} \right) ^{2}\nu-2\,{{ x_3}}^{2}-3\, \left( -{ x_2}+{ m_2} \right) ^{2} \big) {{ a_3}}^{4} \\
 &\qquad\qquad\quad-8\,{{ a_3}}^{3} \big(  \left( { x_3}+{ m_3} \right) \nu-3/4\,{m_3}-1/2\,{ x_3} \big)  \left( -{ x_2}+{ m_2} \right) { a_2} \\
 &\qquad\qquad\quad+4\, \bigg(  \left( { m_2}+{ x_3}+{ m_3}-{ x_2} \right)  \left( -{ m_2}+{ x_3}+{ m_3}+{x_2} \right) \nu \\
&\qquad\qquad\quad\qquad-5/4\,{{ x_3}}^{2}-{ x_3}\,{ m_3}+3/4\, \left( { m_2}-{ m_3}-{ x_2} \right)  \left( { m_2}+{ m_3}-{ x_2} \right)  \bigg) {{ a_3}}^{2}{{
 a_2}}^{2} \\
 &\qquad\qquad\quad+8\,{ a_3}\, \left( -{ x_2}+{ m_2} \right)  \left(  \left( { x_3}+{ m_3} \right) \nu-{ x_3}-3/4\,{ m_3} \right) {{ a_2}}^{3} \\
 &\qquad\qquad\quad-4\, \big(  \big( { x_3}+{ m_3} \big) ^{2}\nu -3/4\,{{ m_3}}^{2}-2\,{ x_3}\,{ m_3}-3/4\,{{ x_3}}^{2} \big) {{ a_2}}^{4} \bigg\} \\
 &\qquad\qquad . \arctan \left( {\frac {t{{ a_2}}^{2}+ \left( -
{ x_2}+{ m_2} \right) { a_2}+{ a_3}\, \left( { x_3}+{ m_3}+{ a_3}\,t \right) }{ \left( { x_3}+{ m_3} \right) { a_2}-{ a_3}\, \left( -{ x_2}+{ m_2}
 \right) }} \right) \\
 &\qquad\qquad-4\,t \left( {{ a_2}}^{2}+{{ a_3}}^{2} \right)  \left( -3/4+\nu \right)  \left( { a_3}\, \left( -{ m_2}+{ x_2} \right) + \left( { x_3}+{ m_3
} \right) { a_2} \right) {{ a_3}}^{2} ~ \bigg]  \\
&\qquad\quad \bigg/ \bigg( { a_3}\, \left( -{ m_2}+{ x_2} \right) + \left( { x_3}+{ m_3} \right) { a_2} \bigg)  \\
& -\frac{x_3}{4\pi \left( -1+\nu \right)} \bigg\{  \big( { a_3}\, \left( -{ m_2}+{ x_2} \right) + \left( { x_3}+{ m_3} \right) { a_2} \big) {{ a_3}}^{3} \bigg( {{ a_3}}^{2}{t}^{2}+2\,t \left( { x_3}+{ m_3} \right) { a_3} \\
&\qquad\qquad\qquad\qquad+{{ a_2}}^{2}{t}^{2}+2\,t \left( -{ x_2}+{ m_2} \right) { a_2}+(x_3+m_3)^2+ \left( x_2-m_2 \right) ^{2} \bigg) \\
&\qquad\qquad . \ln  \bigg[   {t}^{2}+ 2\left( \left( { m_2}-{ x_2} \right) { a_2}+{ a_3}\, \left( { x_3}+{ m_3} \right)  \right) t+(x_2-m_2)^2+(x_3+m_3)^2 \bigg] \\
&\qquad\qquad+ \bigg( -{{ a_3}}^{4}{ x_3}-3\, \left( -{ x_2}+{ m_2} \right) { a_2}\,{{ a_3}}^{3}+{{ a_2}}^{2} \left( 3\,{ m_3}+{ x_3} \right) {{ a_3}}^{2} \\
&\qquad\qquad\qquad\qquad-{{ a_2}}^{3} \left( -{ x_2}+{ m_2} \right) { a_3}+{ m_3}\,{{ a_2}}^{4} \bigg)  \\
&\qquad\qquad\qquad . \big( {{ a_3}}^{2}{t}^{2}+2\,t \left( { x_3}+{ m_3} \right) { a_3} +{{ a_2}}^{2}{t}^{2} \\
&\qquad\qquad\qquad\qquad+2\,t \left( -{ x_2}+{ m_2} \right) { a_2}+(x_3+m_3)^2+ \left( x_2-m_2 \right) ^{2} \big) \\
&\qquad\qquad . \arctan \left[ {\frac {t\,{{a_2}}^{2}+ \left( -{ x_2}+{ m_2} \right) { a_2}+{ a_3}\, \left( { x_3}+{ m_3}+{ a_3}\,t \right) }{ \left( { x_3}+{ m_3} \right) { a_2}-{ a_3}\, \left( -{ x_2}+{ m_2} \right) }} \right] \\
&\qquad\qquad - \bigg( -t{{ a_3}}^{4}{ x_3}+ \left( -3\,t \left( -{ x_2}+{ m_2} \right) { a_2}-{{ x_3}}^{2}-{ x_3}\,{ m_3}- \left( -{ x_2}+{ m_2} \right) ^{2} \right) {{ a_3}}^{3} \\
&\qquad\qquad\qquad+3\,{ a_2}\, \left( t \left( { x_3}+{ m_3} \right) { a_2}-1/3\,{ m_3}\, \left( -{ x_2}+{ m_2} \right) \right) {{ a_3}}^{2} \\
&\qquad\qquad\qquad+{ a_3}{{ a_2}}^{2} \left( t \left( -{ x_2}+{ m_2} \right) { a_2}+{{ x_3}}^{2}+3\,{ x_3}\,{ m_3}+2\,{{ m_3}}^{2}+ \left( -{ x_2}+{ m_2} \right) ^{2} \right)  \\
&\qquad\qquad\qquad-{ m_3}\,{{ a_2}}^{3} \left( -{ x_2}+{ m_2}+{ a_2}\,t \right)  \bigg)  \left( { a_3}\, \left( -{ m_2}+{ x_2} \right) + \left( { x_3}+{ m_3} \right) { a_2} \right)  ~ \bigg\}   \\
&\qquad\left( { a_3}\, \left( -{ m_2}+{ x_2} \right) + \left( { x_3}+{ m_3} \right) { a_2} \right) ^{-1} \left( (x_3+m_3+a_3\,t)^2+(x_2-m_2-a_2\,t)^2\right) ^{-1} \\
I_{23}(t)& = \\
&\frac{1}{4\pi\left( -1+\nu \right)}\, \bigg\{ \frac{1}{4} \big(  \left(  x_3- m_3 \right)  a_2- a_3\, \left( x_2-m_2 \right)  \big)  \left(  a_2- a_3 \right)  \left(  a_2+a_3 \right) \\
&\qquad\qquad . \ln  \left(   t^2+ \left(  \left( 2\, m_2-2\, x_2 \right)  a_2+2\, a_3\, \left( - x_3+m_3 \right)  \right) t+ (x_2-m_2)^2+(x_3-m_3)^2 \right) \\
&\qquad\qquad+ a_2\,{ a_3} \bigg(  \big(  \left(x_3- m_3 \right)  a_2+ a_3\, \left( - x_2+ m_2 \right)  \big) \\
&\qquad\qquad\qquad\qquad . \arctan \left( {\frac {-t{{ a_2}}^{2}+ \left( -{ m_2}+{ x_2} \right) { a_2}-{ a_3}\, \left( { a_3}\,t-{ x_3}+{ m_3} \right) }{ \left( -{ x_3}+{ m_3} \right) { a_2}-{ a_3}\, \left( { x_2}+{ m_2} \right) }} \right) -1/2\,  t \bigg) \, \bigg\}  \\ 
&+\frac{1}{a_3\pi( -1/2+\nu)} \bigg\{ -\frac{1}{2}\, \big( { a_3}\, \left( -{ m_2}+{ x_2} \right) + \left( { x_3}+{ m_3} \right) { a_2} \big) { a_3}\, \\
&\qquad\qquad . \ln  \left[  \left( -{\frac {{ a_2}}{{
a_3}}}+{\frac {{ a_2}\,{ x_3}+{ a_2}\,{ m_3}+{ a_3}\,{ x_2}-{ a_3}\,{ m_2}}{{ a_3}\, \left( { x_3}+{ m_3}+{ a_3}\,t \right) }} \right) ^{2}+1 \right] \\
&\qquad\qquad-
 \left( { a_3}\, \left( -{ m_2}+{ x_2} \right) + \left( { x_3}+{ m_3} \right) { a_2} \right) { a_2}\, \\
&\qquad\qquad . \arctan \left[ {\frac {t{{ a_2}}^{2}+ \left( -{ x_2}+{
 m_2} \right) { a_2}+{ a_3}\, \left( { x_3}+{ m_3}+{ a_3}\,t \right) }{ \left( { x_3}+{ m_3} \right) { a_2}-{ a_3}\, \left( -{ x_2}+{ m_2} \right) }}
 \right] \\
&\qquad\qquad + \big( { a_3}\, \left( -{ m_2}+{ x_2} \right) + \left( { x_3}+{ m_3} \right) { a_2} \big) { a_3}\,\ln  \left[ {\frac {{ a_3}\, \left( -{ m_2}+{
 x_2} \right) + \left( { x_3}+{ m_3} \right) { a_2}}{{ x_3}+{ m_3}+{ a_3}\,t}} \right] \\
&\qquad\qquad+ \left( { a_3}\, \left( -{ m_2}+{ x_2} \right) + \left( { x_3}+{ m_3} \right) { a_2} \right) { a_2}\,\arctan \left( {\frac {{ a_3}}{{ a_2}}} \right) \\
&\qquad\qquad+   \left( { x_3}+{ m_3}+{
a_3}\,t \right) \arctan \left( {\frac {-{ x_2}+{ m_2}+{ a_2}\,t}{{ x_3}+{ m_3}+{ a_3}\,t}} \right) \, \bigg\} \\ 
&+\frac{-3+4\nu}{ 4\pi \left( -1+\nu \right)}  \bigg\{  \bigg(  \left( -1/4\,{ x_2}+1/4\,{ m_2} \right) {{ a_3}}^{3} -3/4\, \left( { x_3}+1/3\,{ m_3} \right) { a_2}\,{{ a_3}}^{2}\\
&\qquad\qquad\qquad\qquad-1/4\,{{ a_2}}^{2} \left( -{ x_2}+{ m_2} \right) { a_3}-1/4\,{{ a_2}}^{3} \left( { x_3}-{ m_3} \right)  \bigg) \\
&\qquad\qquad . \ln  \bigg[   {t}^{2}+ \left(  \left( 2\,{ m_2}-2\,{ x_2} \right) { a_2}+2\,{ a_3}\, \left( { x_3}+{ m_3} \right)  \right) t+(x_2-m_2)^2+(x_3-m_3)^2\bigg] \\
 &\qquad\qquad + \bigg(  \left( {{ a_3}}^{2}{ x_3}+ \left( -{ x_2}+{ m_2} \right) { a_2}\,{ a_3}-{{
a_2}}^{2}{ m_3} \right) \\
&\qquad\qquad\qquad . \arctan \left[ {\frac {t{{ a_2}}^{2}+ \left( -{ x_2}+{ m_2} \right) { a_2}+{ a_3}\, \left( { x_3}+{ m_3}+{ a_3}\,t \right) }{ \left(
{ x_3}+{ m_3} \right) { a_2}-{ a_3}\, \left( -{ x_2}+{ m_2} \right) }} \right] +1/2\,{ a_2}\,t   \bigg) { a_3} \, \bigg\}  \\ 
 &-\frac{1}{4\pi( -1+\nu )}{x_3}\, \bigg\{ { a_2}\, \bigg( {{ a_3}}^{2}{t}^{2}+2\,t \left( { x_3}+{ m_3} \right) { a_3}+{{ a_2}}^{2}{t}^{2}+2\,t \left( -{ x_2}+{ m_2} \right) { 
 a_2} \\
&\qquad\qquad\qquad\qquad\qquad +{{ m_3}}^{2}+2\,{ x_3}\,{ m_3}+{{ x_3}}^{2}+ \left( -{ x_2}+{ m_2} \right) ^{2} \bigg) {{ a_3}}^{2} \\
&\qquad\qquad . \ln  \bigg[  {t}^{2}+ \left(  \left( 2\,{ m_2}-2\,{ x_2} \right) { a_2}+2\,{ a_3}\, \left( { x_3}+{ m_3} \right)  \right) t+(x_2-m_2)^2+(x_3+m_3)^2 \bigg] \\
 &+ { a_3}\left( { a_2}-{ a_3} \right)  \left( { a_2}+{ a_3} \right)  \bigg( {{ a_3}}^{2}{t}^{2}+2
\,t \left( { x_3}+{ m_3} \right) { a_3} +{{ a_2}}^{2}{t}^{2}\\
&\qquad\qquad\qquad+2\,t \left( -{ x_2}+{ m_2} \right) { a_2}+{{ m_3}}^{2}+2\,{ x_3}\,{ m_3}+{{ x_3}}^{2}+
 \left( -{ x_2}+{ m_2} \right) ^{2} \bigg) \, \\
 &\qquad\qquad . \arctan \left[ {\frac {t{{ a_2}}^{2}+ \left( -{ x_2}+{ m_2} \right) { a_2}+{ a_3}\, \left( { x_3}+{
m_3}+{ a_3}\,t \right) }{ \left( { x_3}+{ m_3} \right) { a_2}-{ a_3}\, \left( -{ x_2}+{ m_2} \right) }} \right] \\
&-t \left( -{ x_2}+{ m_2} \right) {{ a_3}}^{4}+ \left( 3\,t \left( { x_3}+1/3\,{ m_3} \right) { a_2}-{ m_3}\, \left( -{ x_2}+{ m_2} \right)  \right) {{ a_3}}^{3} \\
&\qquad\qquad+2\,{ a_2}\, \left( 3/2\,t \left( -{x_2}+{ m_2} \right) { a_2}+1/2\,{{ m_3}}^{2}+3/2\,{ x_3}\,{ m_3}+{{ x_3}}^{2}+ \left( -{ x_2}+{ m_2} \right) ^{2} \right) {{ a_3}}^{2} \\
&\qquad\qquad -{{ a_2}}^{2}\left(t \left( 3\,{ m_3}+{ x_3} \right) { a_2}+{ m_3}\, \left( -{ x_2}+{ m_2} \right)  \right) { a_3}-{ m_3}\,{{ a_2}}^{3} \left( { x_3}+{ m_3} \right) \, \bigg\}  \\
&\qquad \bigg/ \bigg( (x_3+m_3+a_3t)^2+(x_2-m_2-a_2t)^2\bigg) \\
I_{32}(t)& = \\
&\frac{1}{4\pi\left( -1+\nu \right)}\, \bigg\{ \frac{1}{4} \big(  \left(  x_3- m_3 \right)  a_2- a_3\, \left( x_2-m_2 \right)  \big)  \left(  a_2- a_3 \right)  \left(  a_2+a_3 \right) \\
&\qquad\qquad . \ln  \left(   t^2+ \left(  \left( 2\, m_2-2\, x_2 \right)  a_2+2\, a_3\, \left( - x_3+m_3 \right)  \right) t+ (x_2-m_2)^2+(x_3-m_3)^2 \right) \\
&\qquad\qquad+ a_2\,{ a_3} \bigg(  \big(  \left(x_3- m_3 \right)  a_2+ a_3\, \left( - x_2+ m_2 \right)  \big) \\
&\qquad\qquad\qquad\qquad . \arctan \left( {\frac {-t{{ a_2}}^{2}+ \left( -{ m_2}+{ x_2} \right) { a_2}-{ a_3}\, \left( { a_3}\,t-{ x_3}+{ m_3} \right) }{ \left( -{ x_3}+{ m_3} \right) { a_2}-{ a_3}\, \left( { x_2}+{ m_2} \right) }} \right) -1/2\,  t \bigg) \, \bigg\}  \\ 
&+\frac{1}{a_3\pi( -1/2+\nu)} \bigg\{ -\frac{1}{2}\, \big( { a_3}\, \left( -{ m_2}+{ x_2} \right) + \left( { x_3}+{ m_3} \right) { a_2} \big) { a_3}\, \\
&\qquad\qquad . \ln  \left[  \left( -{\frac {{ a_2}}{{
a_3}}}+{\frac {{ a_2}\,{ x_3}+{ a_2}\,{ m_3}+{ a_3}\,{ x_2}-{ a_3}\,{ m_2}}{{ a_3}\, \left( { x_3}+{ m_3}+{ a_3}\,t \right) }} \right) ^{2}+1 \right] \\
&\qquad\qquad-
 \left( { a_3}\, \left( -{ m_2}+{ x_2} \right) + \left( { x_3}+{ m_3} \right) { a_2} \right) { a_2}\, \\
&\qquad\qquad . \arctan \left[ {\frac {t{{ a_2}}^{2}+ \left( -{ x_2}+{
 m_2} \right) { a_2}+{ a_3}\, \left( { x_3}+{ m_3}+{ a_3}\,t \right) }{ \left( { x_3}+{ m_3} \right) { a_2}-{ a_3}\, \left( -{ x_2}+{ m_2} \right) }}
 \right] \\
&\qquad\qquad + \big( { a_3}\, \left( -{ m_2}+{ x_2} \right) + \left( { x_3}+{ m_3} \right) { a_2} \big) { a_3}\,\ln  \left[ {\frac {{ a_3}\, \left( -{ m_2}+{
 x_2} \right) + \left( { x_3}+{ m_3} \right) { a_2}}{{ x_3}+{ m_3}+{ a_3}\,t}} \right] \\
&\qquad\qquad+ \left( { a_3}\, \left( -{ m_2}+{ x_2} \right) + \left( { x_3}+{ m_3} \right) { a_2} \right) { a_2}\,\arctan \left( {\frac {{ a_3}}{{ a_2}}} \right) \\
&\qquad\qquad+   \left( { x_3}+{ m_3}+{
a_3}\,t \right) \arctan \left( {\frac {-{ x_2}+{ m_2}+{ a_2}\,t}{{ x_3}+{ m_3}+{ a_3}\,t}} \right) \, \bigg\} \\ 
&+\frac{-3+4\nu}{ 4\pi \left( -1+\nu \right)}  \bigg\{  \bigg(  \left( -1/4\,{ x_2}+1/4\,{ m_2} \right) {{ a_3}}^{3} -3/4\, \left( { x_3}+1/3\,{ m_3} \right) { a_2}\,{{ a_3}}^{2}\\
&\qquad\qquad\qquad\qquad-1/4\,{{ a_2}}^{2} \left( -{ x_2}+{ m_2} \right) { a_3}-1/4\,{{ a_2}}^{3} \left( { x_3}-{ m_3} \right)  \bigg) \\
&\qquad\qquad . \ln  \bigg[   {t}^{2}+ \left(  \left( 2\,{ m_2}-2\,{ x_2} \right) { a_2}+2\,{ a_3}\, \left( { x_3}+{ m_3} \right)  \right) t+(x_2-m_2)^2+(x_3-m_3)^2\bigg] \\
 &\qquad\qquad + \bigg(  \left( {{ a_3}}^{2}{ x_3}+ \left( -{ x_2}+{ m_2} \right) { a_2}\,{ a_3}-{{
a_2}}^{2}{ m_3} \right) \\
&\qquad\qquad\qquad . \arctan \left[ {\frac {t{{ a_2}}^{2}+ \left( -{ x_2}+{ m_2} \right) { a_2}+{ a_3}\, \left( { x_3}+{ m_3}+{ a_3}\,t \right) }{ \left(
{ x_3}+{ m_3} \right) { a_2}-{ a_3}\, \left( -{ x_2}+{ m_2} \right) }} \right] +1/2\,{ a_2}\,t   \bigg) { a_3} \, \bigg\}  \\ 
 &-\frac{1}{4\pi( -1+\nu )}{x_3}\, \bigg\{ { a_2}\, \bigg( {{ a_3}}^{2}{t}^{2}+2\,t \left( { x_3}+{ m_3} \right) { a_3}+{{ a_2}}^{2}{t}^{2}+2\,t \left( -{ x_2}+{ m_2} \right) { 
 a_2} \\
&\qquad\qquad\qquad\qquad\qquad +{{ m_3}}^{2}+2\,{ x_3}\,{ m_3}+{{ x_3}}^{2}+ \left( -{ x_2}+{ m_2} \right) ^{2} \bigg) {{ a_3}}^{2} \\
&\qquad\qquad . \ln  \bigg[  {t}^{2}+ \left(  \left( 2\,{ m_2}-2\,{ x_2} \right) { a_2}+2\,{ a_3}\, \left( { x_3}+{ m_3} \right)  \right) t+(x_2-m_2)^2+(x_3+m_3)^2 \bigg] \\
 &+ { a_3}\left( { a_2}-{ a_3} \right)  \left( { a_2}+{ a_3} \right)  \bigg( {{ a_3}}^{2}{t}^{2}+2
\,t \left( { x_3}+{ m_3} \right) { a_3} +{{ a_2}}^{2}{t}^{2}\\
&\qquad\qquad\qquad+2\,t \left( -{ x_2}+{ m_2} \right) { a_2}+{{ m_3}}^{2}+2\,{ x_3}\,{ m_3}+{{ x_3}}^{2}+
 \left( -{ x_2}+{ m_2} \right) ^{2} \bigg) \, \\
 &\qquad\qquad . \arctan \left[ {\frac {t{{ a_2}}^{2}+ \left( -{ x_2}+{ m_2} \right) { a_2}+{ a_3}\, \left( { x_3}+{
m_3}+{ a_3}\,t \right) }{ \left( { x_3}+{ m_3} \right) { a_2}-{ a_3}\, \left( -{ x_2}+{ m_2} \right) }} \right] \\
&-t \left( -{ x_2}+{ m_2} \right) {{ a_3}}^{4}+ \left( 3\,t \left( { x_3}+1/3\,{ m_3} \right) { a_2}-{ m_3}\, \left( -{ x_2}+{ m_2} \right)  \right) {{ a_3}}^{3} \\
&\qquad\qquad+2\,{ a_2}\, \left( 3/2\,t \left( -{x_2}+{ m_2} \right) { a_2}+1/2\,{{ m_3}}^{2}+3/2\,{ x_3}\,{ m_3}+{{ x_3}}^{2}+ \left( -{ x_2}+{ m_2} \right) ^{2} \right) {{ a_3}}^{2} \\
&\qquad\qquad -{{ a_2}}^{2}\left(t \left( 3\,{ m_3}+{ x_3} \right) { a_2}+{ m_3}\, \left( -{ x_2}+{ m_2} \right)  \right) { a_3}-{ m_3}\,{{ a_2}}^{3} \left( { x_3}+{ m_3} \right) \, \bigg\}  \\
&\qquad \bigg/ \bigg( (x_3+m_3+a_3t)^2+(x_2-m_2-a_2t)^2\bigg)
\end{align*}
And finally,
\normalsize
\begin{align*}
I_{33}(t)& = \\
&-\frac{3-4\nu}{8\pi (1-\nu) }  \bigg\{ \frac{1}{2} \left( t  - \left( {x_2}-{m_2} \right) {a_2}-{a_3}\, \left( {x_3}-{m_3}           \right)  \right) \\
&\qquad\qquad . \ln  \bigg[t^{2}+ \left(  \left( 2\,{m_2}-2\,{x_2} \right) {a_2}+2\,{a_3}\, \left( -{x_3}+{           m_3} \right)  \right) t+({x_2}-m_2)^{2}+(x_3-m_3)^2\bigg]  \\ 
&+ \left(  \left( {x_3}-{m_3} \right) {a_2}+{a_3}\, \left( -{x_2}+{m_2} \right)  \right) \arctan \bigg[ {\frac {-t+ \left( -{m_2}+{x_2} \right) {a_2}-{a_3}\, \left(-{x_3}+{m_3} \right) }{ \left( -{x_3}+{m_3} \right) {a_2}-{a_3}\, \left( -{x_2}+{m_2} \right) }} \bigg] ~ \bigg\} \\ 
&+\frac{1}{8\pi(1-\nu)} \bigg\{ -{ a_3}\,{ a_2}\, \left(  \left( { x_3}-{ m_3} \right) { a_2}+{ a_3}\, \left( -{ x_2}+{ m_2} \right)  \right) \\
&\qquad\qquad. \ln  \bigg[  {t}^{2}+ \left(  \left( 2\,{ m_2}-2\,{ x_2} \right) { a_2}+2\,{ a_3}\, \left( -{ x_3}+{ m_3} \right)  \right) t+ (x_2-m_2)^{2}+(x_3-m_3)^{2} \bigg] \\
&+ \left( { a_2}-{ a_3} \right)  \left( { a_2}+{ a_3} \right)  \left(  \left( {
 x_3}-{ m_3} \right) { a_2}+{ a_3}\, \left( -{ x_2}+{ m_2} \right)  \right) \\
 &\qquad\qquad . \arctan \left[ {\frac {-t+ \left( -{ m_2}+{ x_2} \right) { a_2}-{
 a_3}\, \left(-{ x_3}+{ m_3} \right) }{ \left( -{ x_3}+{ m_3} \right) { a_2}-{ a_3}\, \left( -{ x_2}+{ m_2} \right) }} \right] -t{{ a_2}}^{2} ~ \bigg\} \\ 
&+\frac{8\nu^2-12\nu+5}{8\pi (-1+\nu )} \bigg\{  \frac{1}{2}\left( t a_2^2+ \left( m_2-x_2 \right)  a_2+a_3\, \left(  x_3+ m_3+ a_3\,t \right)  \right) \\
&\qquad\qquad . \ln  \bigg[  t^2+ \left(  \left( 2\, m_2-2\, x_2 \right)  a_2+2\, a_3\, \left(
 x_3+ m_3 \right)  \right) t+ (x_2-m_2)^2+ (x_3-m_3)^2 \bigg]  \\
 &+ \left(  \left(  x_3+ m_3 \right)  a_2- a_3\, \left( - x_2+ m_2 \right)  \right) \\
 &\qquad\qquad . \arctan \left[ \frac{ t+ \left( - x_2+ m_2 \right)  a_2+ a_3\,
 \left(  x_3+ m_3 \right) }{ \left(  x_3+ m_3 \right)  a_2- a_3\, \left( - x_2+ m_2 \right)  }\right] \,  \bigg\} \\ 
&+\frac{1}{8\pi (-1+\nu) } \bigg\{ 4\,{ a_3}\, \big(  \left( { x_3}+{ m_3} \right) { a_2}-{ a_3}\, \left( -{ x_2}+{ m_2} \right)  \big)  \\
&\qquad . \left(  \left(  \left( { x_3}+{ m_3} \right) \nu-{ x_3}-3/4\,{ m_3} \right) {{ a_2}}^{2}- \left( -3/4+\nu \right) { a_3}\, \left( -{ x_2}+{ m_2} \right) { a_2}-1/4\,{{ a_3}}^{2}{ x_3} \right) \\
&\qquad\qquad . \ln  \bigg[  {t}^{2}+ 2t\left(  \left( { m_2}-{ x_2} \right) { a_2}+{ a_3}\, \left( { x_3}+{ m_3} \right) \right)+(x_2-m_2)^{2}+(x_3-m_3)^2\bigg] \\
&+ \bigg[-4\, \bigg(  \left( { m_2}+{ x_3}+{ m_3}-{ x_2} \right)  \left( -{ m_2}+{ x_3}+{ m_3}+{ x_2} \right) \nu \\
&\qquad\qquad\qquad-5/4\,{{ x_3}}^{2}-{ x_3}\,{ m_3}+3/4\, \left( { m_2}-{ m_3}-{ x_2} \right) \left( { m_2}+{ m_3}-{ x_2} \right)  \bigg) {{ a_3}}^{2}{{ a_2}}^{2} \\
&\qquad+ \left( 4\, \left( { x_3}+{ m_3} \right) ^{2}\nu-3\,{{ m_3}}^{2}-8\,{ x_3}\,{ m_3}-3\,{{ x_3}}^{2} \right) {{ a_2}}^{4} \\
&\qquad+{{ a_3}}^{3} \left( 8 \left( { x_3}+{ m_3} \right) \nu-6\,{ m_3}-4\,{ x_3} \right)  \left( -{ x_2}+{ m_2} \right) { a_2} \\
&\qquad-8\,{ a_3}\, \left( -{ x_2}+{ m_2} \right)  \left(  \left( { x_3}+{ m_3} \right) \nu-{ x_3}-3/4\,{ m_3} \right) {{ a_2}}^{3} \\
&\qquad+\left( -4\, \left( -{ x_2}+{ m_2} \right) ^{2}\nu+{{ x_3}}^{2}+3\, \left( -{ x_2}+{ m_2} \right) ^{2} \right) {{ a_3}}^{4} \bigg] \\
&\qquad\qquad . \arctan \left[ {\frac {t{{ a_2}}^{2}+ \left( -{ x_2}+{ m_2} \right) { a_2}+{ a_3}\, \left( { x_3}+{ m_3}+{ a_3}\,t \right) }{ \left( { x_3}+{ m_3} \right) { a_2}-{ a_3}\, \left( -{ x_2}+{ m_2} \right) }} \right] \\
&+\left( 3-4\nu \right) t \left(  \left( { x_3}+{ m_3} \right) { a_2}-{ a_3}\, \left( -{ x_2}+{ m_2} \right)  \right)   {{ a_2}}^{2}  ~ \bigg\} \\
&\qquad\qquad  \bigg/ \bigg(  \left( { x_3}+{ m_3} \right) { a_2}-{ a_3}\, \left( -{ x_2}+{ m_2 } \right)  \bigg)   \\ 
&+\frac{1}{8\pi (-1+\nu)} \bigg\{ 4\,{ a_3}\, \left(  \left( { x_3}+{ m_3} \right) { a_2}-{ a_3}\, \left( -{ x_2}+{ m_2} \right)  \right)  \\
&\qquad . \left(  \left(  \left( { x_3}+{ m_3}
 \right) \nu-{ x_3}-3/4\,{ m_3} \right) {{ a_2}}^{2}- \left( -3/4+\nu \right) { a_3}\, \left( -{ x_2}+{ m_2} \right) { a_2}-1/4\,{{ a_3}}^{2}{ x_3} \right) \\
&\qquad . \ln  \left(   {t}^{2}+ 2\left(  \left( { m_2}-{ x_2} \right) { a_2}+{ a_3}\, \left( { x_3}+{ m_3} \right) \right) t+(x_2-m_2)^{2}+(x_3-m_3^{2} \right)  \\
&\qquad+ \bigg[  \left( 4\, \left( { x_3}+{ m_3} \right) ^{2}\nu-3\,{{ m_3}}^{2}-8\,{ x_3}\,{ m_3}-3\,{{ x_3}}^{2} \right) {{ a_2}}^{4} \\
&\qquad\qquad-8\,{ a_3}\, \left( -{ x_2}+{ m_2} \right)  \left(  \left( { x_3}+{
 m_3} \right) \nu-{ x_3}-3/4\,{ m_3} \right) {{ a_2}}^{3} \\
 &\qquad\qquad-4\, \bigg(  \left( { m_2}+{ x_3}+{ m_3}-{ x_2} \right)  \left( -{ m_2}+{ x_3}+{ m_3}+{ x_2} \right) \nu-5/4\,{{ x_3}}^{2}-{ x_3}\,{ m_3} \\
 &\qquad\qquad\qquad\qquad+3/4\, \left( { m_2}-{ m_3}-{ x_2} \right)  \left( { m_2}+{ m_3}-{ x_2} \right)  \bigg) {{ a_3}}^{2}{{ a_2}}^{2} \\
&\qquad\qquad+8\,{{ a_3}}^{3} \left(  \left( { x_3}+{ m_3} \right) \nu-3/4\,{ m_3}-1/2\,{ x_3} \right)  \left( -{ x_2}+{ m_2} \right) { a_2} \\
&\qquad\qquad+2\, \left( -2\, \left( -
{ x_2}+{ m_2} \right) ^{2}\nu+{{ x_3}}^{2}+3/2\, \left( -{ x_2}+{ m_2} \right) ^{2} \right) {{ a_3}}^{4} \bigg] \\
&\qquad\qquad\qquad . \arctan \left[ {\frac {t{{ a_2}}^{2}+ \left( -{ x_2}+{ m_2} \right) { a_2}+{ a_3}\, \left( { x_3}+{ m_3}+{ a_3}\,t \right) }{ \left( { x_3}+{ m_3} \right) { a_2}-{ a_3}\, \left( -{ x_2}+{ m_2} \right) }} \right] \\
&\qquad-4\, \left( -3/4+\nu \right) t \left(  \left( { x_3}+{ m_3} \right) { a_2}-{ a_3}\, \left( -{ x_2}+{ m_2} \right)  \right)   {{ a_2}}^{2} ~ \bigg\}  \\
& \qquad \qquad \bigg/ \bigg(  \left( { x_3}+{ m_3} \right) { a_2}-{ a_3}\, \left( -{ x_2}+{ m_2} \right)  \bigg) ~.
 \end{align*}
The equations for $I_{23}$ and $I_{32}$ are numerically unstable in the limit $a_3=0$. For this special case, given that $||\textbf{a}||=1$, I take the analytical limit for $a_2\rightarrow1$ and $a_3\rightarrow0$. The codes to evaluate these equations, as well as any other solution presented in this manuscript, are available in an online repository (see the Data and Resources section).

\end{document}